\documentclass[aps,prd,superscriptaddress,12pt,nofootinbib]{revtex4}

\usepackage{amsfonts}
\usepackage{amsmath,epsfig,bm}
\usepackage{graphicx}
\usepackage{color}

\usepackage[setpagesize=false]{hyperref}

\newcommand{\be}{\begin{equation}}
\newcommand{\ee}{\end{equation}}
\newcommand{\ba}{\begin{eqnarray}}
\newcommand{\ea}{\end{eqnarray}}
\newcommand{\bd}{\begin{displaymath}}
\newcommand{\ed}{\end{displaymath}}

\def\thalf{{\textstyle{\frac{1}{2}}}}

\def\oneth{{\textstyle{\frac{1}{3}}}}

\newcommand{\mb}[1]{\mathbf{#1}}

\begin{document}

\title{Thermal Conductivity and Chiral Critical Point \\ in Heavy Ion Collisions}

\author{Joseph I. Kapusta}
\affiliation{School of Physics \& Astronomy, University of Minnesota, Minneapolis, MN 55455,USA}
\author{Juan M. Torres-Rincon}
\affiliation{Departamento de Fisica Teorica I, Universidad Complutense de Madrid, 28040 Madrid, Spain}

\date{December 9, 2012}

\begin{abstract}
\begin{description}
\item[Background:]  Quantum Chromodynamics is expected to have a phase
transition in the same static universality class as the 3D Ising model and
the liquid-gas phase transition.  The properties of the equation of state,
the transport coefficients, and especially the location of the critical
point are under intense theoretical investigation.  Some experiments are
underway, and many more are planned, at high energy heavy ion
accelerators.
\item[Purpose:]   Develop a model of the thermal conductivity, which diverges
at the critical point, and use it to study the impact of hydrodynamic
fluctuations on observables in high energy heavy ion collisions.
\item[Methods:]  We apply mode coupling theory, together with a previously
developed model of the free energy that incorporates the critical
exponents and amplitudes, to construct a model of the thermal conductivity
in the vicinity of the critical point.  The effect of the thermal
conductivity on correlation functions in heavy ion collisions is studied
in a boost invariant hydrodynamic model via fluctuations, or noise.
\item[Results:]  We find that the closer a thermodynamic trajectory comes to
the critical point the greater is the magnitude of the fluctuations in
thermodynamic variables and in the 2-particle correlation functions in
momentum space.
\item[Conclusions:] It may be possible to discern the existence of a
critical point, its location, and thermodynamic and transport properties
near to it in heavy ion collisions using the methods developed here.
\end{description}
\end{abstract}

\maketitle
\newpage

\section{Introduction}
\label{sec:introduction}

It has now been firmly established, via lattice calculations, that QCD with its physical quark masses does not exhibit a phase transition at finite temperature and zero baryon density but only a very rapid crossover from quark-gluon plasma-like behavior to hadronic gas-like behavior \cite{Aoki2006a}-\cite{Bazavov2012}.  However, since the up and down quark masses are so small, and chiral symmetry is nearly exact, there are reasons to suspect that there is a curve of first order phase transition in the temperature $T$ versus baryon chemical potential $\mu$ plane, terminating at a critical point at $T_c > 0$ and $\mu_c > 0$.   The existence of such a critical point has been found in various effective field theory models, such as the Nambu--Jona-Lasinio model \cite{asakawa89}-\cite{scavenius01}, a composite operator model \cite{barducci}, a random matrix model \cite{halasz98}, a linear $\sigma$ model \cite{scavenius01}, an effective potential model \cite{hatta02}, and a 
hadronic bootstrap model \cite{antoniou02}.  Lattice QCD has yet to confirm or deny the existence of a critical point.  The reason is that inclusion of a chemical potential does not allow for straight-forward Monte Carlo samplings of the field configurations.  For reviews see Refs. \cite{stephanov} and \cite{MohantyQM}.    

High energy heavy ion collisions may provide experimental evidence for a critical point and provide information on the behavior of the equation of state in its vicinity.  Relevant to this are low energy runs at the Relativistic Heavy Ion Collider (RHIC), and in the future at the Facility for Antiproton and Ion Research (FAIR), at the SPS Heavy Ion and Neutrino Experiment (SHINE), and at Nuclotron-based Ion Collider Facility (NICA).  Critical points are characterized by large fluctuations.  This led to the suggestion to study fluctuations in conserved quantities, such as electric charge, baryon number, and strangeness on an event-by-event basis  \cite{Shuryak1998,Hatta2003}.  The effect is proportional to the spatial size of the domain or correlated region, which is probably rather small due to the finite size and lifetime in heavy ion collisions \cite{small}.  Therefore, it was suggested to measure higher moments to search for non-Gaussian behavior \cite{nongaussian}. However, at present there is no 
experimental evidence for anomalous fluctuations of this kind \cite{STAR2010}.

A crucial issue is whether the critical point can ever be reached in a heavy ion collision.  Colliding nuclei is necessary to reach high baryon density, but at the same time it creates entropy.  If the initial entropy per baryon is much larger than that at the critical point, then the expanding matter will never pass close to it, even under the assumption of an ideal adiabatic expansion, since entropy can only increase with time, not decrease.  The problem is similar to that of trying to create superheavy nuclei by colliding nuclei: too much entropy is created.

In this paper we construct a semi-realistic model for the thermal conductivity due to an assumed critical point in the QCD phase diagram.  At a critical point the thermal conductivity diverges, as does the shear viscosity.  However, the critical exponent for the shear viscosity is much smaller than that for the thermal conductivity, with the implication that the influence of the divergence of the shear viscosity is confined to a very narrow window in temperature, probably too small to have any effect on the matter produced in a heavy ion collision.  In contrast, the temperature window for the enhancement of the thermal conductivity is much wider.  These transport coefficients, along with the bulk viscosity, control the strength of hydrodynamic fluctuations in heavy ion collisions \cite{Kapusta2012}.  In particular, the strength of the hydrodynamic fluctuations due to thermal conductivity $\lambda$ are quantified by the correlation function
\be
\langle f(x_1) f(x_2) \rangle = 2 \lambda \left( \frac{nT}{sw}\right)^2 \delta(x_1 - x_2) \ ,
\label{bfluctuations}
\ee
where $f(x)$ is a dimensionless fluctuation, or noise term, that appears in the hydrodynamic equations.  Reference \cite{Kapusta2012} applied the relativistic theory of hydrodynamic fluctuations to heavy ion collisions, with a specific example worked out for matter created with zero average baryon density.\footnote{Earlier studies for extracting the shear viscosity in heavy ion collisions were done by \cite{Gavin-Aziz} and for the bulk viscosity by \cite{Dobado}.} In this paper we will focus on thermal conductivity and ignore viscosities for simplicity of exposition.  This allows us to study the influence of a critical point on the produced matter in a controlled and quantitative manner, although the model we use is not realistic enough for direct comparison with experiment.  The Landau-Lifshitz definition of flow velocity is used in this paper since it naturally extends to baryon-free matter, whereas the Eckart definition becomes ambiguous in that limit; see, for example, \cite{KapustaGale, Kapusta2012}.  
Due to this rational choice of reference frame the behavior of the shear and bulk viscosities near the critical point only affect the background solution of the hydrodynamic equations for which, in this exploratory study, we use the perfect fluid solutions.  Even greater effects may be anticipated when all of these effects are incorporated self-consistently.

Reference \cite{Kapusta2010} constructed the free energy in the vicinity of the critical point that incorporated both the critical exponents and critical amplitudes.  Using this free energy, the Landau theory of fluctuations was applied to estimate the probability of fluctuations away from the equilibrium state and were found to be very large.  However, the Landau theory considers baryon number fluctuations in a finite volume in contact with a heat reservoir at fixed temperature, which does not adequately represent the space-time evolution of matter in a heavy ion collision.  Dynamical simulations of spinodal decomposition, or phase separation, were done in \cite{Bower:2001fq} and \cite{Randrup2009}, but without the incorporation of intrinsic hydrodynamical fluctuations or noise.  In this article we assume that the entropy created in the collision is too high to allow any trajectories in the $T-\mu$ plane to pass through the coexistence region.  Rather, following the earlier remarks on entropy in heavy
ion collisions, we assume that the trajectories always pass to the left of the critical point, never entering the coexistence region or crossing the line of first order phase transition.  This is a more conservative scenario for heavy ion experiments. 

Hydrodynamic fluctuations, or noise, may be crucial for studying the effects of a critical point in heavy ion collisions.  As an analogy, Ref.  \cite{Eggers} performed a theoretical study of the breakup of liquid nanojets with the conclusion that ``noise is the driving force behind pinching, speeding up the breakup to make surface tension irrelevant".  Similar conclusions were reached in Ref. \cite{Kang}, which studied the breakup of liquid nanobridges.

The outline of the article is as follows.  In Sec. 2 we construct a semi-realistic model for the critical enhancement of the thermal conductivity.  This is based on the mode-coupling theory which has been successfully applied to the liquid-gas phase transition in a variety of ordinary atomic systems.  In Sec. 3 we discuss the equation of state to be used in conjunction with the thermal conductivity to model the expansion phase of heavy ion collisions.  In Sec. 4 we implement these ideas in a boost-invariant hydrodynamic model; although not the most relevant fluid dynamic model for studying the critical point, it allows us to gain insight and intuition, and to discover the magnitude of the effect on observables in Sec. 5.  We summarize and conclude in Sec. 6.

It should be acknowledged that there are other sources of fluctuations in heavy ion collisions, such as initial state fluctuations, fluctuations induced by jets and other high momentum-transfer processes, and fluctuations during hadronization in the final state.  These were surveyed in Ref. \cite{Kapusta2012}.  Fluctuations caused by passage near a critical point should be characterized by a strong beam energy dependence.  Fluctuations due to jets should not be relevant at the energies where the critical point might be reached.

\section{Thermal Conductivity}
\label{sec:conductivity}

In ordinary materials, thermal conductivity is typically inferred from measurements of a diffusion constant rather than measured directly.  There are several kinds of diffusion constants depending on the experimental conditions.  The baryon diffusion constant $D_B$ is defined by the diffusion equation
\be
\frac{\partial X}{\partial t} =  D_B \nabla^2 X \ ,
\ee
where $X$ is the departure from the equilibrium baryon density $n$ or baryon chemical potential $\mu$ at fixed temperature $T$.  The fully relativistic expression for this diffusion constant is
\be
D_B = \frac{\lambda T}{(\partial n/\partial \mu)_T} \left(\frac{n}{w}\right)^2 \ ,
\ee
where $w$ is the enthalpy density and $\lambda$ is the thermal conductivity (dimension of energy-squared).  The partial derivative in this expression is just the baryon number susceptibility
\be
\chi_B = (\partial n/\partial \mu)_T \ .
\ee
The isothermal compressibility $\kappa_T$ is
\be
\kappa_T^{-1} = n \left(\frac{\partial P}{\partial n}\right)_T \ ,
\ee
where $P$ is the pressure.  These are simply related by $\chi_B = n^2 \kappa_T$.  

The diffusion equation for heat is
 \be
\frac{\partial T}{\partial t} =  D_T \nabla^2 T \ ,
\ee
which is carried out at constant pressure.  This diffusion constant can be derived from first order relativistic viscous fluid dynamics to be
\be
D_T = \frac{\lambda}{c_P} \ ,
\ee
where $c_P = T(\partial s/\partial T)_P$ is the heat capacity per unit volume at constant pressure.  

For liquids and gases near their critical point it is easier to measure $D_T$ than $D_B$.  As is well-known, the thermal conductivity diverges at such a critical point.  It is conventional and useful to separate the thermal conductivity into a smooth background piece $\lambda^b$ and a piece $\Delta \lambda$ which diverges at the critical point and which goes to zero away from the critical point so that $\lambda = \lambda^b + \Delta \lambda$.  Exactly how this is done is not unique and is considered a bit of an art.  Then $\Delta \lambda = c_P \Delta D_T$.  

We would like to know the thermal conductivity not only in the asymptotic critical region but in the non-critical region too.  For this one must go beyond the renormalization group, which is only valid asymptotically close to the critical point.  A rather successful approach to this problem is to use mode-coupling theory for the dynamics of critical fluctuations in fluids extended away from the critical region \cite{Fixman1962}-\cite{Zwanzig1972}.  The basic idea is the recognition that in a fluid the slow modes are the diffusive modes of heat and viscosity while the sound mode is considered a fast mode.  The approach is quite general, and has been developed for fluids near and above the critical temperature in the universality class of the liquid-gas phase transition.  Only the essential points will be reviewed and summarized here \cite{Sengers1995}.

The part $\Delta D_T$ due to the critical enhancement takes the form
\be
\Delta D_T = \frac{R_D T}{6\pi\eta\xi} \Omega(q_D\xi) \ .
\ee
Here $\xi$ is the correlation length, $\eta$ is the shear viscosity, $R_D$ is a universal constant approximately equal to 1.05, and $\Omega$ is a crossover function which goes to zero as $q_D \xi$ goes to zero and which goes to 1 as $q_D \xi$ goes to 
infinity. The $q_D$ is a cutoff in wavenumber and is material-dependent\footnote{To be precise, both $\Delta D_T$ and the crossover function $\Omega$ depend on the wavenumber $q$.  In the hydrodynamic limit one takes $q \xi \rightarrow 0$ while $q_D \xi$ remains finite.  Hence we consider $\Omega(q_D\xi)$ a function with a single argument.}.
As the critical point is approached, $\xi \rightarrow \infty$, and the Stokes-Einstein diffusion law is recovered
\be
\Delta D_T \rightarrow \frac{R_D T}{6\pi\eta\xi} \ .
\ee
The smooth background value of $\eta$, without the critical enhancement, is sometimes used because its critical exponent of 0.063 for the reduced temperature \cite{Hao1991}-\cite{Nieuwoudt1989} is much smaller than the critical exponent for the thermal conductivity, and its divergence at the critical point contributes negligibly to the results.

A model for the crossover function $\Omega(q_D \xi)$, which partly accounts for non-asymptotic critical behavior for the thermal conductivity, was presented in Fig. 4 of \cite{Sengers1995}.  The complicated numerical results can be approximated to two-significant digits in the range $0.5 < q_D \xi < \infty$ by
\be
\Omega(x) \approx 0.48 \tanh(0.23 x) + \frac{1.04}{\pi} \arctan (0.65 x) \ .
\ee
The correlation length needs a precise definition.   Reference~\cite{Sengers1995} uses
\be
\xi = \xi_0 \left( \frac{\Delta \chi_B^*}{\Gamma}\right)^{\nu/\gamma} \ ,
\ee
where 
\be
\chi_B^* = \left(\frac{P_c}{n_c^2}\right)\chi_B = \frac{P_c}{n_c^2}\left(\frac{\partial n}{\partial \mu}\right)_T
\ee
 is a dimensionless susceptibility.  Both $\xi_0$ and $\Gamma$ are smoothly and slowly varying functions of $n$ and $T$.  A function
\be
\Delta \chi_B^*(n,T) = \chi_B^*(n,T) - \chi_B^*(n,T_{\rm ref})\frac{T_{\rm ref}}{T}
\label{Sengerchi}
\ee
was used in \cite{Sengers1995} so that the enhancement from criticality goes to zero at some reference temperature $T_{\rm ref}$.  The critical exponents are $\gamma \approx 1.24$ for the isothermal compressibility or baryon number susceptibility, and $\nu \approx 0.63$ for the correlation length.  These are the critical exponents for systems in the same static universality class as the liquid-gas phase transition and the 3D Ising model; it is generally accepted that the QCD phase transition under discussion belongs to the same universality class \cite{Rajagopal2000}.  It has further been shown \cite{Son2004} that this QCD phase transition belongs to the dynamic universality class of model H of Ref. \cite{Hohenberg1977}.

An explicit expression is needed for the correlation length.  Based on  the work in \cite{Kapusta2010} we take the baryon number susceptibility in the critical region to be
\be
\Delta \chi_B^* = \frac{ 5 P_c}{(\delta + 1) f_{\sigma}}
\left[ \left( \frac{\delta - 1}{2 - \gamma} \right)\left(\frac{\Delta n}{n_c} \right) t^\gamma + 5 \delta \, |\eta|^{\delta - 1}\right]^{-1} \ .
\label{Kapustachi}
\ee
This results in the correlation length
\be
\xi(n,T) = \bar{\xi}_0 \left[ \left( \frac{\delta - 1}{2 - \gamma} \right) \left( \frac{\Delta n}{n_c} \right) t^\gamma + 5 \delta \, |\eta|^{\delta - 1} \right]^{-\nu/\gamma} \ .
\ee
Here $t \equiv (T-T_c)/T_c \ge 0$ and $\eta \equiv (n-n_c)/n_c$ (not to be confused with the shear viscosity).  The other critical exponent is $\delta = 4.815$.  The $\Delta n$ is the discontinuity in the baryon density at zero temperature.  It was estimated in \cite{Kapusta2010} as $\Delta n = n_c/3$. All the other constants or slowly varying quantities are absorbed into
\be
\bar{\xi}_0 = \xi_0 \left(\frac{5 P_c}{(\delta + 1) f_{\sigma} \Gamma}\right)^{\nu/\gamma} \ .
\ee
Note that the expression (\ref{Kapustachi}) automatically goes to zero, or becomes very small, far away from the critical point.  Hence there is no absolute need to make a subtraction as in (\ref{Sengerchi}).  Finally, when considering heavy ion collisions later on, it is necessary to know $\Delta \lambda$ for $t$ somewhat less than zero but still outside the phase coexistence region.  To do so we shall replace $t^{\gamma}$ in the previous formulas with $|t|^{\gamma}$, which is a smooth extrapolation outside the domain of the derivation.

Now we come to a discussion of the remaining parameters.  Reference \cite{Sengers1995} made a theoretical fit to the transport properties of carbon dioxide and ethane.  Their fit parameters are: $\xi_0 = 1.5$\AA, $q_D^{-1} = 2.0$\AA, and $\Gamma = 0.0481$ for carbon dioxide, and $\xi_0 = 1.9$\AA, $q_D^{-1} = 1.83$\AA, and $\Gamma = 0.0541$ for ethane.  (Note that light scattering can be used to independently fix the critical amplitude $\xi_0$ in these materials, but such is not possible with the matter produced in high energy heavy ion collisions.)  The first point to notice is that $\xi_0 \approx q_D^{-1}$ for both substances.  They are smaller than the average particle separation of $n_c^{-1/3} = 5.39$\AA (carbon dioxide) and $n_c^{-1/3} = 6.23$\AA (ethane), and comparable to the physical dimensions of the molecules.  We will take $q_D \xi_0 = 1$ and assume that the cutoff is $q_D = \pi T_0$ where $T_0$ is the crossover temperature at zero chemical potential; see
the next section.  For the other parameters we choose $\Gamma = 0.05$ and from \cite{Kapusta2010} $f_{\sigma} = 5 P_c$.  Finally, with $T_0 = 170$ MeV (for example) we get $\xi_0 = 0.37$ fm and $\bar{\xi}_0 = 0.69$ fm.
   
There are various theoretical approaches for calculating the background thermal conductivity $\lambda^b$.  A calculation in QCD to lowest order in $\alpha_s$ but to all orders in $\ln \alpha_s$ gives \cite{AMY2003}
\be
D_B = \frac{0.165}{\alpha_s^2 \ln(0.497/\alpha_s)}\frac{1}{T} \ .
\ee
This is for two quark flavors; similar results were obtained for three flavors.  This result requires that $\alpha_s \ll 1$ which is valid only at asymptotically high temperatures.  No calculations have been done with nonzero chemical potentials with this accuracy.  Earlier, Ref. \cite{Hosoya1985} estimated the thermal conductivity with a chemical potential for two flavors as
\be
\lambda^b \approx \frac{0.08}{\alpha_s^2 \ln(1/\alpha_s)}\left(\frac{w}{n}\right)^2
\ee
using the relaxation time approximation to the Boltzmann equation.  

The relaxation time approximation is especially useful for calculating the background thermal conductivity in the hadronic phase.  From \cite{Gavin1985} and \cite{Hosoya1985}
\be
\lambda^b =\frac{1}{3T^2} \sum_i (2s_i+1) {\rm e}^{b_i \mu/T}\int \frac{d^3p}{(2\pi)^3} \tau_i(p) \left(\frac{p}{\varepsilon_i}\right)^2
\left(\varepsilon_i - b_i \frac{w}{n}\right)^2 {\rm e}^{-\varepsilon_i/T} \ ,
\ee
where the sum is over all particles $i$ with baryon number $b_i$, spin $s_i$, and single particle energy $\varepsilon_i = \sqrt{p^2+m_i^2}$.  The $\tau_i(p)$ is the momentum-dependent relaxation time.  In the limit that the baryon chemical potential goes to zero this results in
\be
\lambda^b \left( \frac{nT}{w}\right)^2 = \frac{1}{3} \sum_i (2s_i+1) b_i^2 \int \frac{d^3p}{(2\pi)^3} \tau_i(p) \left(\frac{p}{\varepsilon_i}\right)^2
{\rm e}^{-\varepsilon_i/T} \ .
\ee
Hence the baryon number fluctuations, as reflected by (\ref{bfluctuations}), remain even when the average baryon density is zero.  Note that in this limit only baryons contribute to the thermal conductivity.  This limit will be a good approximation when the enthalpy per baryon is much greater than the nucleon mass. 

It should be remarked that in some special situations the number of particles, in particular pions, is held fixed due to relatively slow particle 
creation and annihilation reactions.  In these situations the net baryon density is replaced by the total particle density, for example $n_{\pi}$.  See, for example, \cite{Gavin1985} and \cite{Prakash1993}.  In the context of unitarized chiral perturbation theory the thermal and electrical conductivities, which are related through the Wiedemann-Franz law, have been
calculated in \cite{torresphd}. In \cite{FernandezFraile:2005ka} the electrical conductivity has been related to the total photon yield $\omega dN_{\gamma}/d^3p$ in the limit $p_T \rightarrow 0$.  However, due to our higher temperatures --where pion number is not conserved-- we will not follow that path here and consider net baryon density.

\section{Equation of State}
\label{sec:EOS}

We need a smooth background equation of state for two reasons.  First, we need the isobaric specific heat in order to convert the thermal diffusion constant to the thermal conductivity in the critical region.  Second, we need an equation of state to solve the background hydrodynamic equations to find the space-time evolution of the independent thermodynamic variables $T$ and $\mu$.

The isobaric specific heat diverges at the critical point.  It obeys the thermodynamic relation
\be
c_P - c_V = \frac{\chi_B T}{n^2} \left(\frac{dP}{dT}\right)_n^2
= \chi_B T \left[  \left(\frac{\partial \mu}{\partial T}\right)_n + \frac{s}{n}\right]^2 \ .
\label{cP-V}
\ee
The isochoric specific heat has critical exponent $\alpha = 0.11$ while the susceptibility has critical exponent $\gamma = 1.24$.  Hence the critical exponent for the isobaric specific heat is $\gamma$ and for the thermal conductivity it is $\gamma - \nu = 0.61$.  This is under the reasonable assumption that the coefficient of $\chi_B$ in the expression for $c_P - c_V$ has a nonzero finite value at the critical point.  It should be noted that use of the equation of state which includes the critical amplitudes and exponents as given in \cite{Kapusta2010} results in the vanishing of $c_P - c_V$.  In particular, the function $\bar{f}_0(t)$ in that paper is only given to first order in $T$ but it is needed to second order to compute $c_P$.  This just means that a smooth background equation of state must be used to calculate the coefficient of $\chi_B$ in (\ref{cP-V}).

In \cite{Kapusta2010} it was assumed that the critical point lies somewhere along a crossover curve parameterized by
\be
\left(\frac{T}{T_0}\right)^2 + \left(\frac{\mu}{\mu_0}\right)^2 = 1 \ ,
\label{Xover}
\ee
with an estimate that $T_0 \approx 170$ MeV and $\mu_0 \approx 1200$ MeV.  The high energy density equation of state was taken to of the form
\be
P = A_4 T^4 + A_2 T^2 \mu^2 + A_0 \mu^4 - C T^2 - B \ .
\label{TsquareEOS}
\ee
This equation of state was used to calculate the critical density, critical entropy density, and so on once $T_c$ was given.  

\begin{figure}
\includegraphics[width=0.8\linewidth]{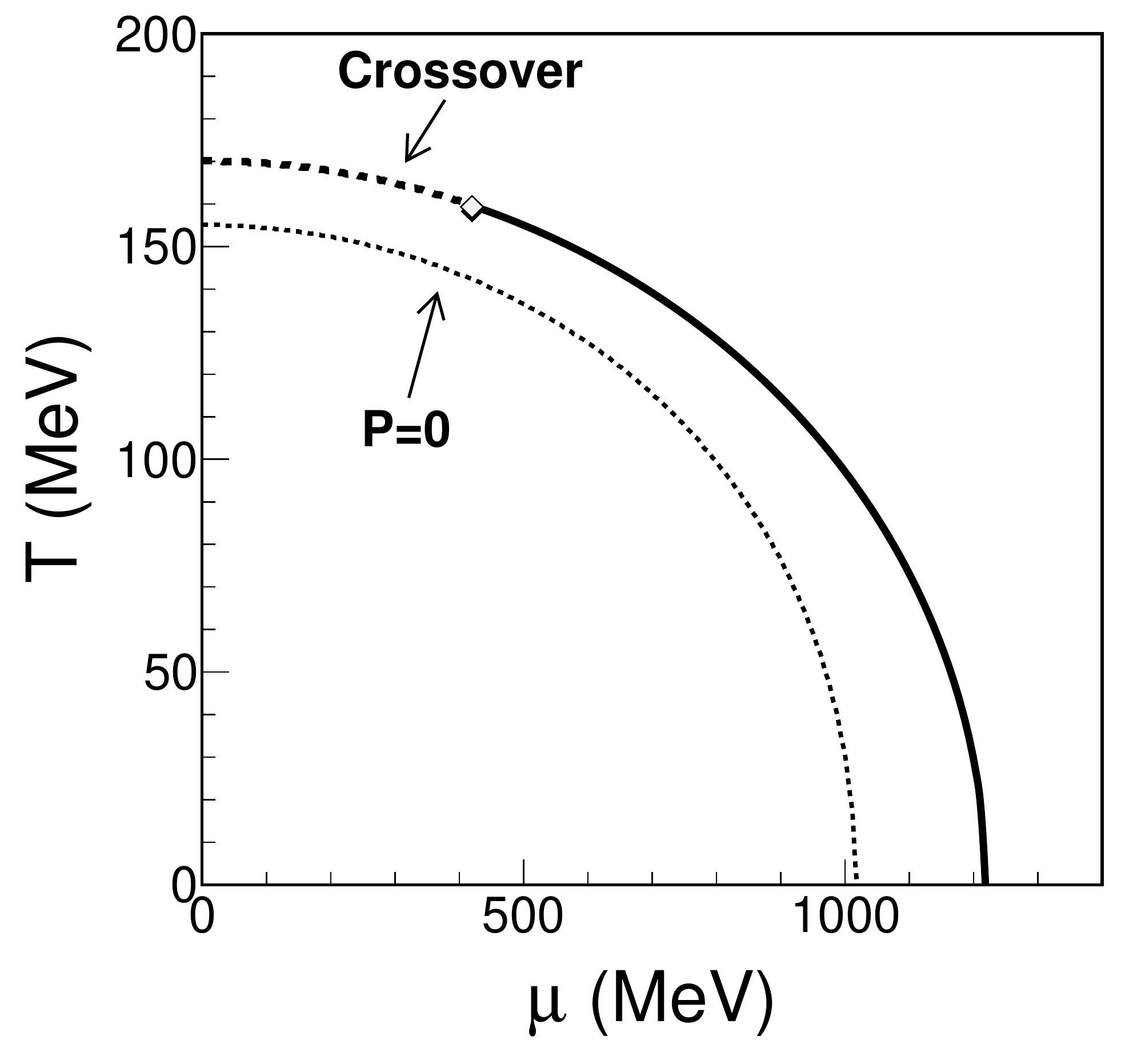}
  \caption{The phase diagram showing the crossover curve and the zero pressure curve.}
  \label{TvsMu}
\end{figure}

If either $P$ or $\epsilon$ is assumed to be constant along the crossover curve then one has the following condition on $\mu_0/T_0$.
\be
A_0 \frac{\mu_0^4}{T_0^4} - A_2 \frac{\mu_0^2}{T_0^2} + A_4 = 0 \ .
\ee
There is also a condition on $C$.  If $P$ is constant along the crossover, as argued in \cite{Kapusta2010} then
\be
C = \mu_0^2\left( A_2  - 2A_0 \frac{\mu_0^2}{T_0^2} \right) \ .
\ee
For definiteness we shall take the coefficients $A_i$ for those of a noninteracting gas of massless gluons and $N_f$ flavors of massless quarks.
\ba
A_4 &=& \frac{\pi^2}{90} \left( 16 + \frac{21N_f}{2} \right) \nonumber \ , \\
A_2 &=& \frac{N_f}{18} \nonumber \ , \\
A_0 &=& \frac{N_f}{324 \pi^2} \ .
\ea
Using $N_f=2$ and $T_0=170$ MeV results in $\mu_0 = 1218.48$ MeV. At the critical point
\be
\frac{w_c}{n_c} = \mu_0 \left( 1 - \frac{T_c^2}{T_0^2}\right)^{-1/2}
\ee
and
\be
\frac{s_c}{n_c} = \left(\frac{\mu_0}{T_0}\right)^2 \frac{T_c}{\mu_c}
\ee
for the both this equation of state and for the parameterization near the critical point constructed in \cite{Kapusta2010}.  Finally, as in \cite{Kapusta2010} we take $B = 0.8 T_0^2$.  The phase diagram is illustrated in Fig. \ref{TvsMu}.

Using  (\ref{TsquareEOS}) the terms appearing in (\ref{cP-V}) are
\be
\left(\frac{\partial \mu}{\partial T}\right)_n  = \frac{-2A_2 T \mu}{A_2 T^2 + 6 A_0 \mu^2} \ ,
\ee
\be
\frac{s}{n} = \frac{T}{\mu} \left(\frac{2 A_4 T^2 + A_2 \mu^2 - C}{A_2 T^2 + 2 A_0 \mu^2}\right) \ .
\ee
These are smooth in the vicinity of the critical point and provide a nonvanishing coefficient of $\chi_B$.  Note that if there was no $T^2$ term in the pressure, $C=0$, then an adiabatic expansion would imply that $T/\mu = $ constant.  In that case the factor
\bd
\left[  \left(\frac{\partial \mu}{\partial T}\right)_n + \frac{s}{n}\right]^2 
\ed
is constant during adiabatic expansion and cooling.

\begin{figure}[top]
\includegraphics[width=0.52\linewidth]{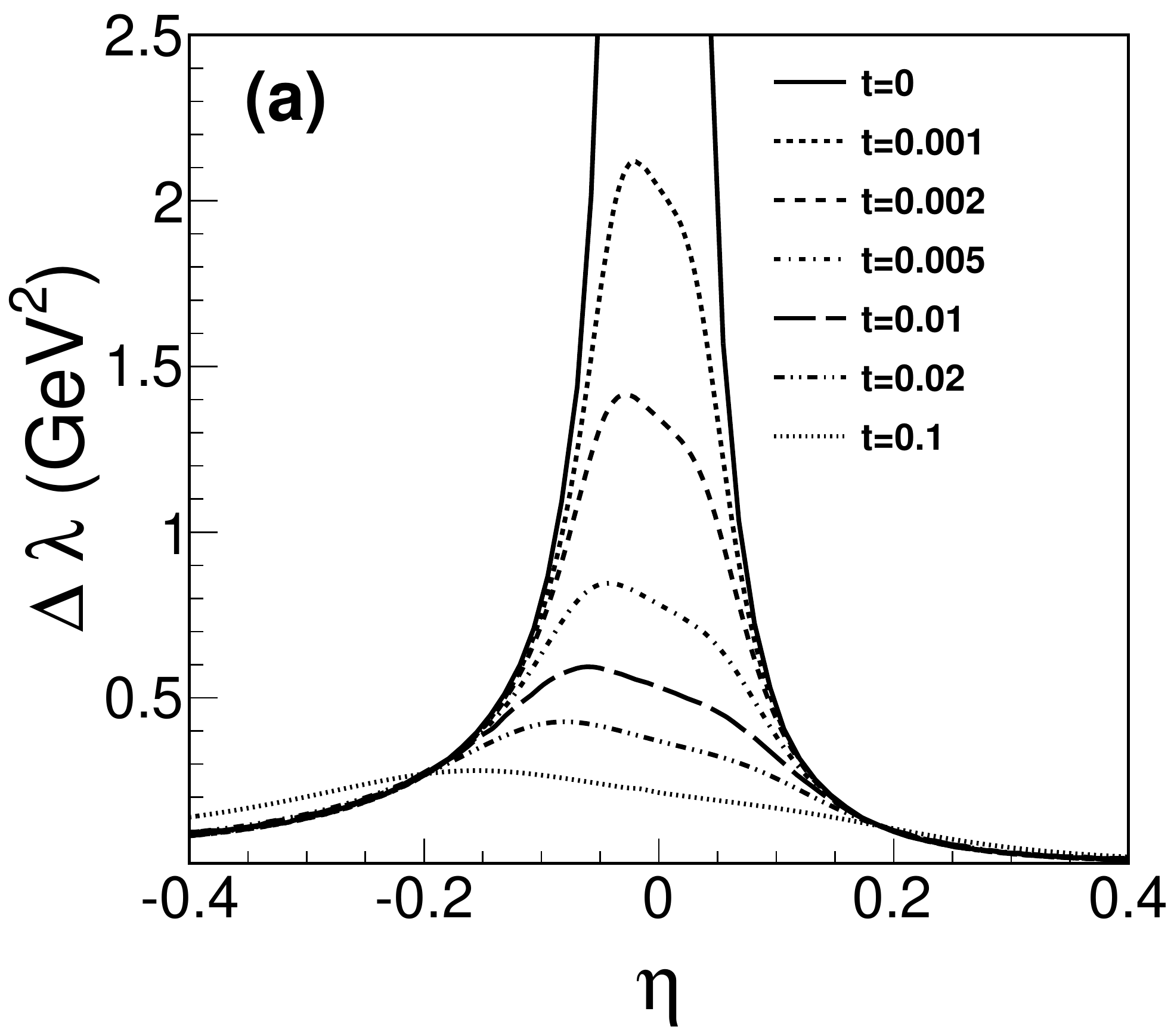}
\includegraphics[width=0.6\linewidth]{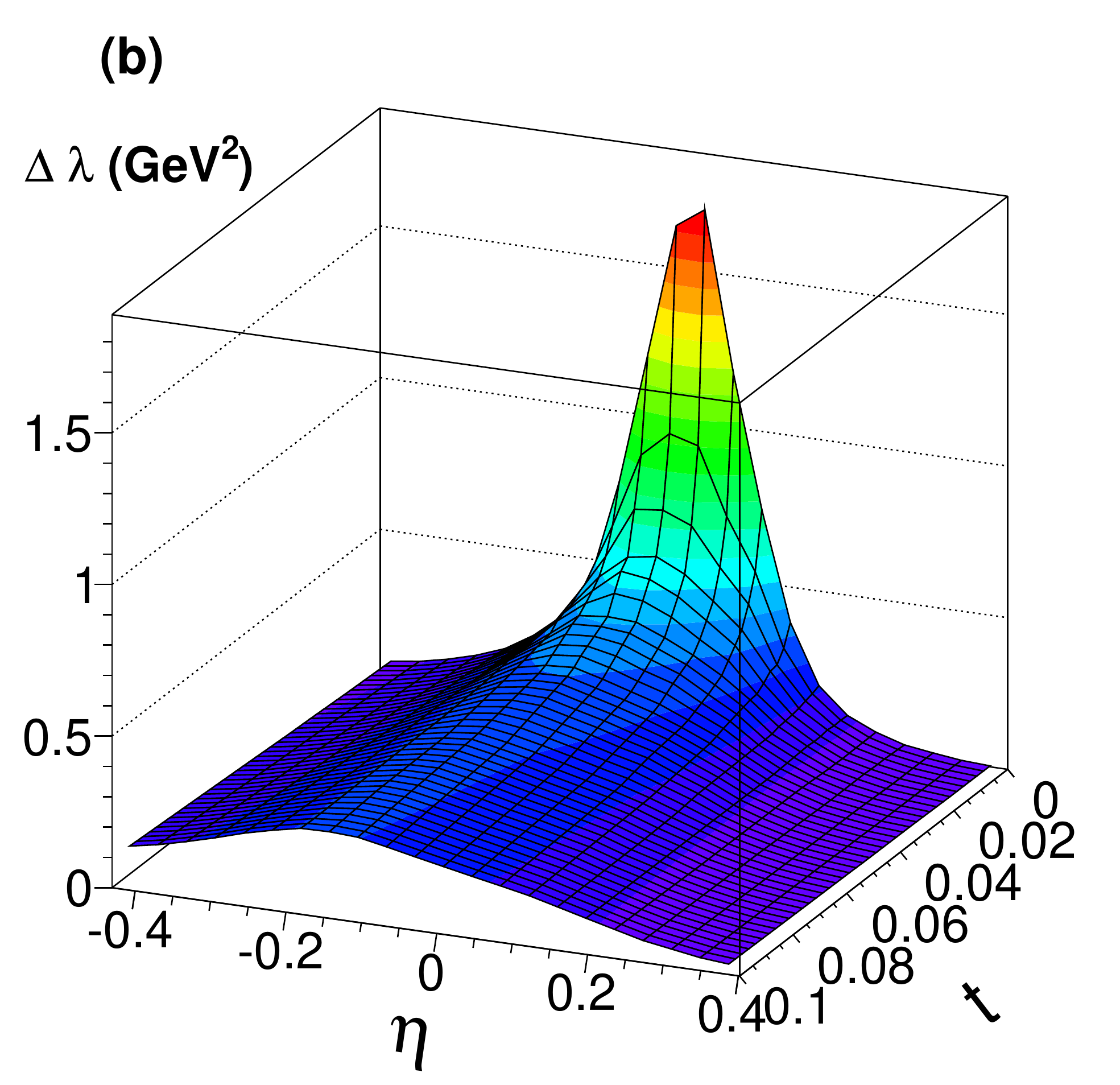}
  \caption{(Color online) Plots of $\Delta \lambda$ using the parameters given in the text.}
  \label{lambdaplots}
\end{figure}

Combining the results from the previous section and this one, we have found a representation for $\Delta \lambda$.  In Fig. \ref{lambdaplots} we plot it as a function of $\eta$ for fixed values of $t$ and in a contour plot as a function of $\eta$ and $t$.  Note that the energy scale is GeV, not tens or hundreds of MeV.  Therefore one should expect significant fluctuations if the trajectory of the expanding matter takes it near the critical point.  In the remaining part of this article we will neglect the background thermal conductivity and use only $\Delta \lambda$.  If the magnitudes of the resulting correlations are observably large, then they would only be greater if $\lambda^b$ was included.

To conclude this section we should point out that one can add higher derivative terms to the dissipative part of the baryon current in the Landau-Lifshitz frame of reference.  This is discussed in Appendix~\ref{app:baryoncurrent}. 

\section{Fluctuations in Boost Invariant Hydrodynamics}
\label{sec:an-example}

In this section we will study the effects of fluctuations on the expanding and cooling matter produced in heavy ion collisions if the trajectory in the $T-\mu$ plane passes near the critical point.  For this purpose we will use the 1+1 dimensional boost-invariant (Bjorken) hydrodynamic model, similar to what was done in Ref.~\cite{Kapusta2012}.  This model is too idealized to make any sort of direct comparison to experimental data.  In addition, this model assumes highly relativistic beam energies, probably much higher than is required to achieve the high baryon densities necessary to probe the critical point.  Nevertheless, it does provide some guidance and intuition before one attempts to study the problem with much more sophisticated and numerically intensive 3+1 dimensional viscous fluid dynamics.  Since the method and equations are so similar to those in \cite{Kapusta2012} we will leave out some of the details.  The gaps can easily be filled in with a little effort.

\begin{figure}
\includegraphics[width=0.49\linewidth]{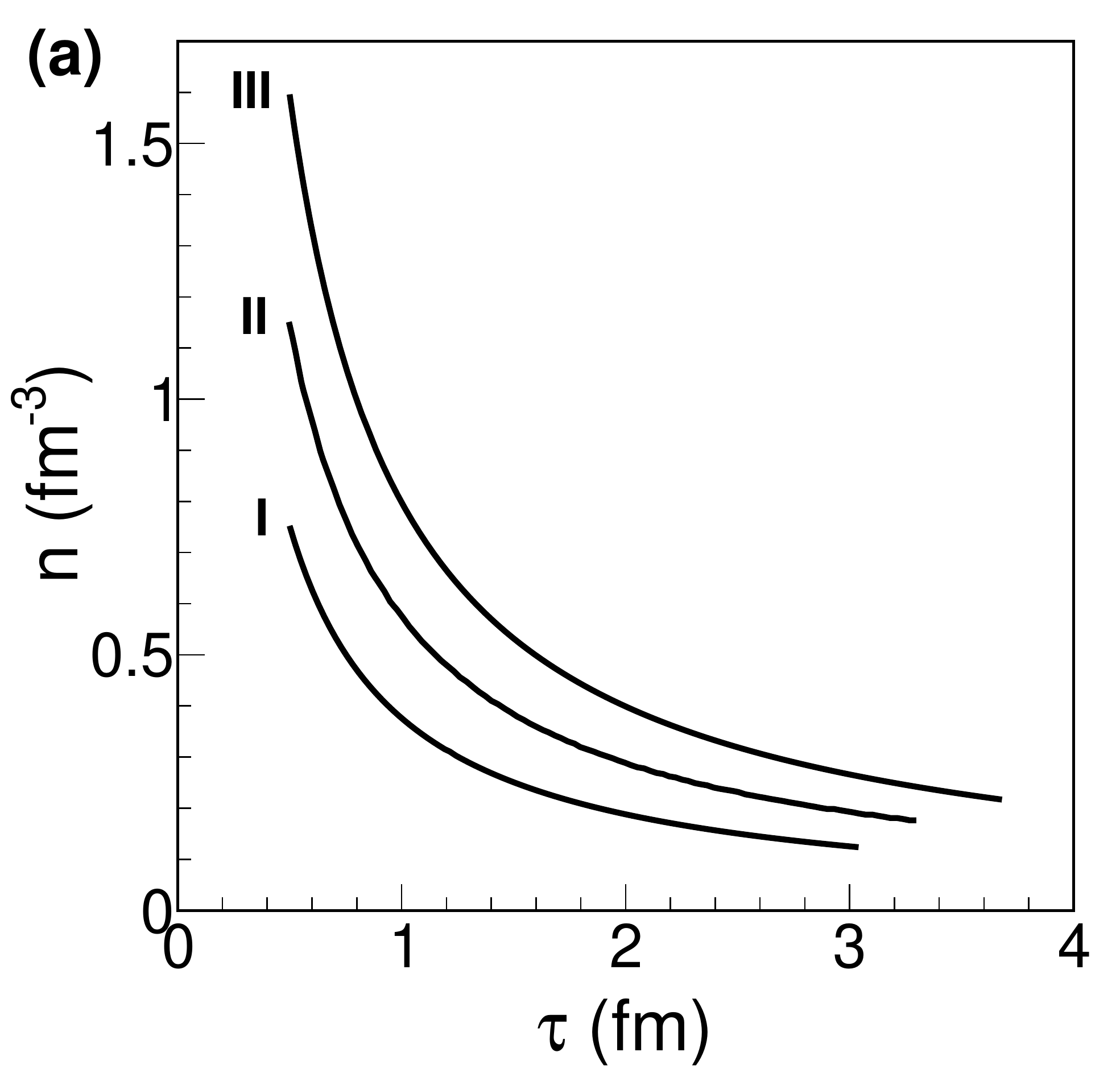}
\includegraphics[width=0.49\linewidth]{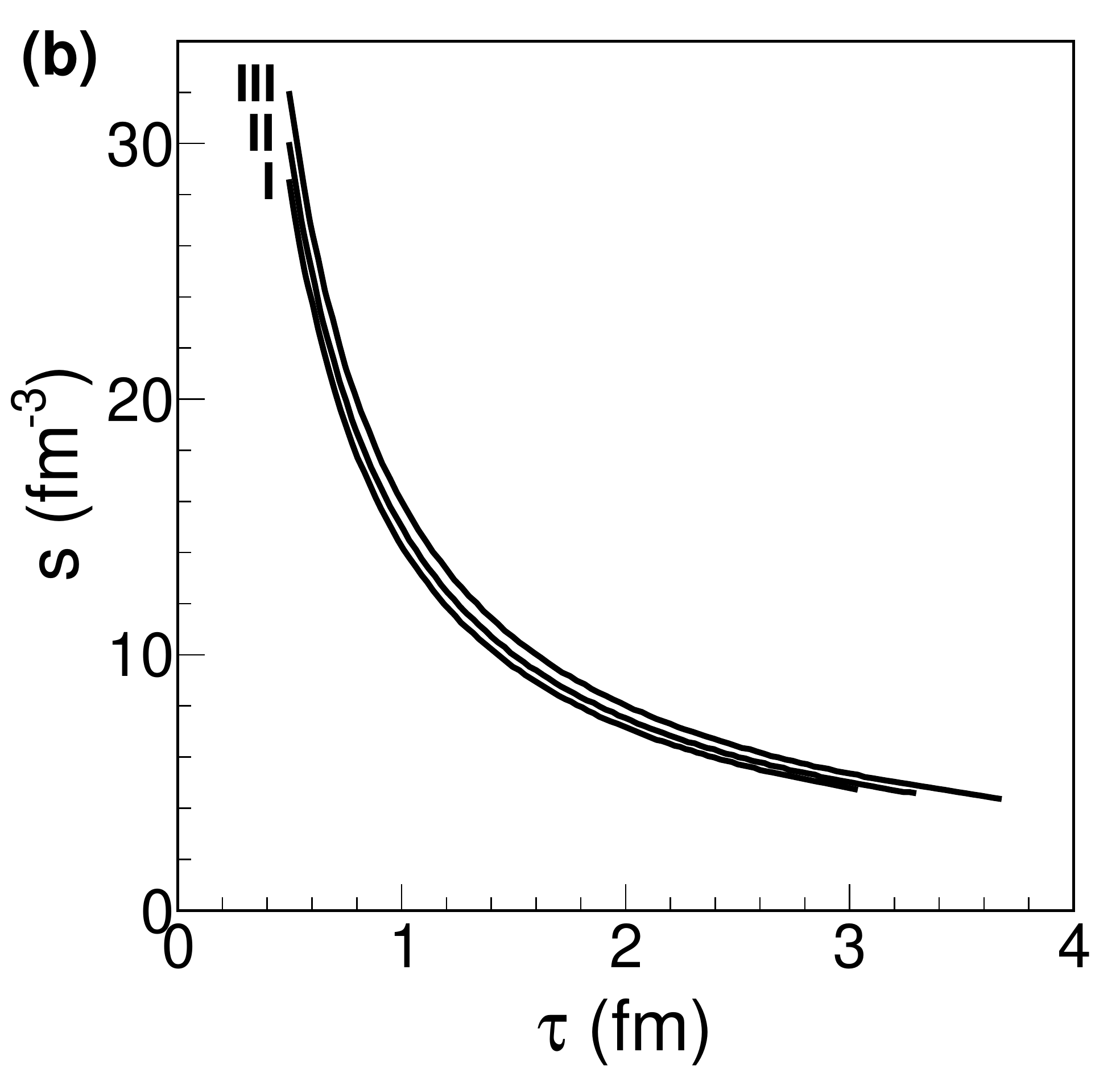}
  \caption{Evolution of baryon density (a) and entropy density (b) as a function of time. They fall as $1/\tau$ so that the entropy per baryon is constant within the
hydrodynamic expansion.}
  \label{fig:scaling}
\end{figure}

\begin{figure}
\includegraphics[width=0.8\linewidth]{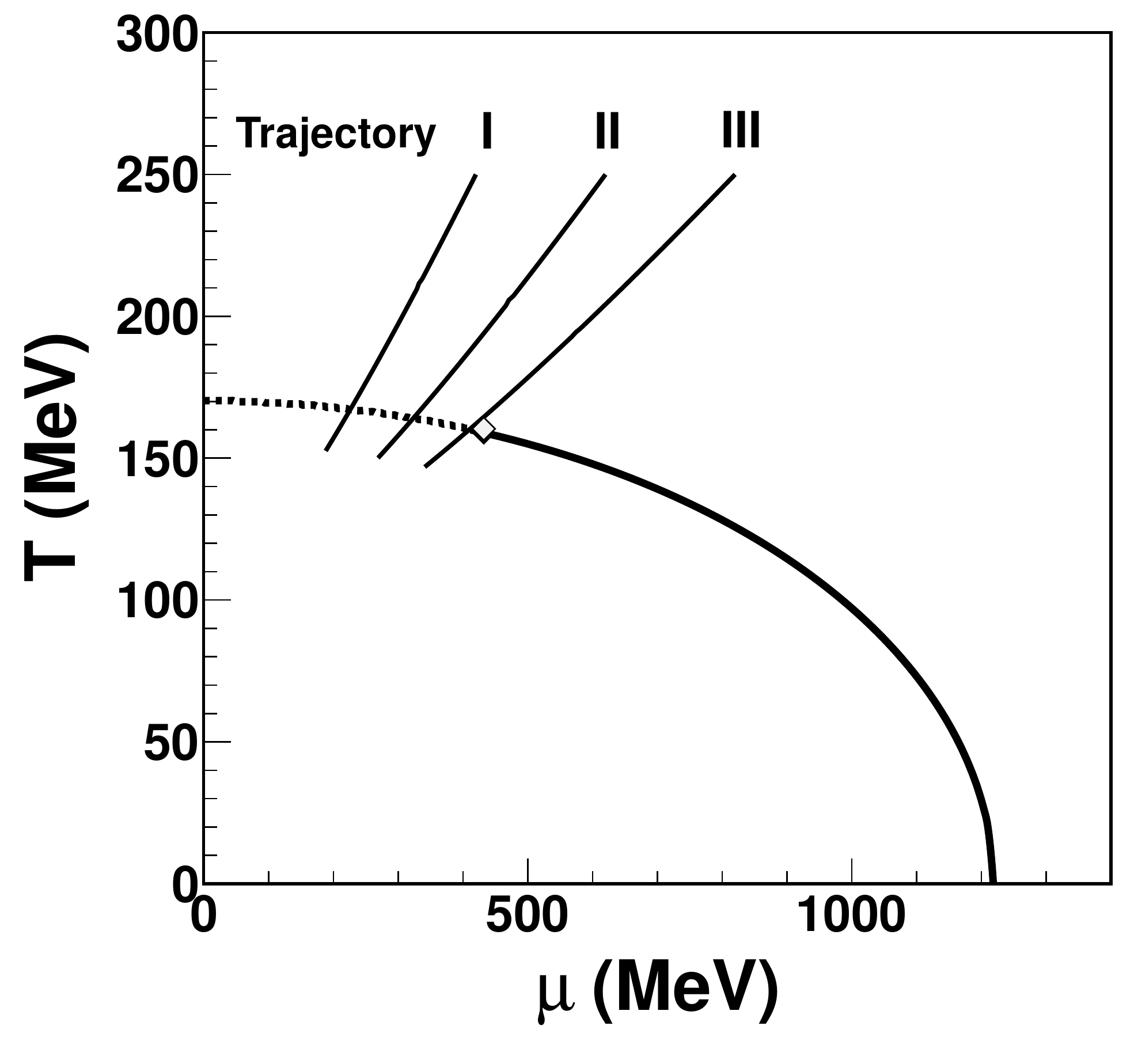}
  \caption{The phase diagram showing the crossover curve and the three trajectories used in the computation.}
  \label{fig:trajectories}
\end{figure}

The energy-momentum tensor in ideal fluid dynamics is
\be
T^{\mu\nu}=wu^{\mu}u^{\nu}-Pg^{\mu\nu} \ .
\ee
The shear and bulk viscosities are ignored to focus on the effects of thermal conductivity.  In the boost-invariant hydrodynamics the flow velocity has the nonvanishing components
\ba
u^0 &=& \cosh(\xi + \omega) \nonumber \ , \\
u^3 &=& \sinh(\xi + \omega) \ . 
\ea
Here $\xi$ is the space-time rapidity (not to be confused with the correlation length) and $\omega(\xi,\tau)$ is a fluctuation that depends on both $\xi$ and the proper time $\tau$.  The baryon current is 
\be
J^{\mu} = n u^{\mu} + I^{\mu} \ ,
\ee
where $I^{\mu}$ is a fluctuation, as described in \cite{Kapusta2012}.  The smooth, background fluid equations lead to the simple equations of motion
\be \label{eq:eomfors}
\frac{ds}{d\tau} + \frac{s}{\tau}=0
\ee
and 
\be 
\label{eq:eomforn}
\frac{dn}{d\tau} + \frac{n}{\tau}=0 \ , 
\ee
independent of the specific equation of state.  The solutions are
\be 
\label{eq:sevol}
s(\tau) = s_i \tau_i/\tau
\ee
and 
\be 
\label{eq:nevol}
n(\tau) = n_i \tau_i/\tau \ ,
\ee
where $s_i$ and $n_i$ are the entropy and baryon densities at some initial time $\tau_i$. Some representative solutions for $s$ and $n$ for later use are represented in Fig. \ref{fig:scaling}.  The initial temperature is $T_i = 250$ MeV, the initial time is $\tau_i = 0.5$ fm/c, and the initial chemical potential was chosen to be $\mu_i =$ 420, 620, and 820 MeV for the three cases.
The adiabatic trajectories, corresponding to these three cases, and are also shown in the phase diagram in Fig. \ref{fig:trajectories}.
Trajectories I and II represent crossover transitions while trajectory III passes very close to the critical end point, which is here chosen to be located at $T_c = 160$ MeV and $\mu_c = 411.74$ MeV.  The entropy per baryon at the critical point is 19.96, while for trajectories I, II and III it is 37.98, 26.08 and 20.06, respectively.  The time evolution is stopped when the zero pressure curve is reached, which is at $\tau_f = $ 3.04, 3.30, and 3.68 fm/c, respectively.  In reality, matching to a full hadronic equation of state should be done, but we don't do it for this illustrative example. 

The full equations are linearized in the fluctuations, such as $\delta n$, and these in turn are linear functionals of $I^{\mu}$.  The nonvanishing components are
\ba
I^0 &=& s(\tau) f(\xi,\tau) \sinh \xi \nonumber \ , \\
I^3 &=& s(\tau) f(\xi,\tau) \cosh \xi
\ea  
on account of the condition that $u_{\mu} I^{\mu}=0$. Notice that $f$ is dimensionless as the entropy density has been factorized out for convenience. The average value of $f(\xi,\tau)$ is zero and the correlation with itself was given in Eq. (\ref{bfluctuations}).  The linearized equations are
\be
\label{eom1a} \tau \frac{\partial \delta \epsilon}{\partial \tau} + \delta w + w \frac{\partial \omega}{\partial \xi} = 0 \ ,
\ee
\be
\label{eom2a} \tau \frac{\partial \delta n}{\partial \tau} + \delta n + n \frac{\partial \omega}{\partial \xi} + s \frac{\partial f}{\partial \xi} = 0 \ ,
\ee
\be
\label{eom3a} \tau \frac{\partial (w\omega)}{\partial \tau} + 2 w\omega + \frac{\partial \delta P}{\partial \xi} = 0 \ .
\ee
Here $n$ and $w$ are the smooth background solutions which depend only on $\tau$.  Note that the noise drives the baryon fluctuations, and if the average baryon density is zero there is no coupling to $\omega$ and so there is only one equation (\ref{eom2a}) to solve.

It is useful to identify the independent variables in Eqs. (\ref{eom1a})-(\ref{eom3a}) as $\delta s$, $\delta n$, and $\omega$.  Then they take the form
\be
\label{eom1b} \tau \frac{\partial \delta s}{\partial \tau} + \delta s + s \frac{\partial \omega}{\partial \xi} - \frac{\mu s}{T} \frac{\partial f}{\partial \xi} = 0 \ ,
\ee
\be
\label{eom2b} \tau \frac{\partial \delta n}{\partial \tau} + \delta n + n \frac{\partial \omega}{\partial \xi} + s \frac{\partial f}{\partial \xi} = 0 \ ,
\ee
\be
\label{eom3b} \tau \frac{\partial \omega}{\partial \tau} + (1-v_{\sigma}^2)\omega + \frac{v_n^2T}{w}\frac{\partial \delta s}{\partial \xi} 
+ \frac{v_s^2\mu}{w}\frac{\partial \delta n}{\partial \xi} = 0 \ .
\ee
Here $v_{\sigma}^2$ is the physical, adiabatic speed of sound squared, while $v_n^2 \equiv (\partial P/\partial \epsilon)_n$ and $v_s^2 \equiv (\partial P/\partial \epsilon)_s$.  They are related by 
\be
v_{\sigma}^2 = \frac{Tsv_n^2 + \mu nv_s^2}{w} \ .
\ee
See Appendix~\ref{app:speedsound}.  Finally, it is convenient to use dimensionless independent variables.  The linearized equations of motion are written in terms of $\delta s/s$, $\delta n/s$ and $\omega$:
\be
\label{eom1c} \tau \frac{\partial}{\partial \tau} \left( \frac{\delta s}{s} \right) + \frac{\partial \omega}{\partial \xi} - \frac{\mu}{T} \frac{\partial f}{\partial \xi} = 0 \ ,
\ee
\be
\label{eom2c} \tau \frac{\partial}{\partial \tau} \left( \frac{\delta n}{s} \right)  + \frac{n}{s} \frac{\partial \omega}{\partial \xi} + \frac{\partial f}{\partial \xi} = 0 \ ,
\ee
\be
\label{eom3c} \tau \frac{\partial \omega}{\partial \tau} + (1-v_{\sigma}^2)\omega + \frac{v_n^2Ts}{w}\frac{\partial}{\partial \xi} \left( \frac{\delta s}{s} \right)
+ \frac{v_s^2\mu s}{w}\frac{\partial}{\partial \xi} \left( \frac{\delta n}{s} \right) = 0 \ .
\ee

In Fourier space
\be
\tilde{X}(k,\tau) = \int_{-\infty}^{\infty} d\xi {\rm e}^{-ik\xi} X(\xi,\tau)
\ee
for any variable $X$.  Then equations (\ref{eom1c})-(\ref{eom3c}) can be expressed as a Langevin equation:
\be 
\tau \frac{\partial \mb{\tilde{\psi}}}{\partial \tau} + \mb{D} \mb{\tilde{\psi}} + \tilde{f}\mb{\tilde{n}} = 0 \ , 
\ee
where the vector $\tilde{\mb{\psi}}$, the drift matrix $\mb{D}$ and the stochastic noise term $\mb{\tilde{n}}$ read:
\be \mb{\tilde{\psi}} = \left( 
\begin{array}{c}
\displaystyle{\frac{\delta \tilde{s}}{s}} \\[7pt]
\displaystyle{\frac{\delta \tilde{n}}{s}} \\
\displaystyle{ \tilde{\omega}}
\end{array}
\right) \ , \quad 
\mb{D} =  \left(
\begin{array}{ccc}
 0  & 0 & ik   \\
  0 & 0 & ik\displaystyle{\frac{n}{s}}  \\
 ik v_n^2 \displaystyle{\frac{Ts}{w}}   & ik v_s^2 \displaystyle{\frac{\mu  s}{w}}  &1 -v^2_\sigma  
\end{array} \right) \ , \quad
\mb{\tilde{n}} = ik \left( 
\begin{array}{c}
  - \displaystyle{\frac{\mu}{T}} \\ 1 \\
  0
\end{array}
\right) \ . 
\ee
The homogeneous equation is solved by the evolution operator $\tilde{\mb{U}} (k;\tau,\tau')$ that is calculated as
\be 
\tilde{{\mb U}} (k; \tau, \tau') = {\mathcal T} \exp \left\{ - \int_{\tau'}^{\tau} \frac{d\tau''}{\tau''} {\mb D} (k,\tau'') \right\}  \ ,
\ee 
where ${\mathcal T}$ is the time-ordering operator.  Once that is known, the solution to the inhomogeneous equation can be expressed as
\be
\tilde{\mb{\psi}}(k,\tau) = - \int_{\tau_i}^{\tau} \frac{d\tau'}{\tau'}\tilde{{\mb U}} (k; \tau, \tau')  \mb{\tilde{n}}(k,\tau') \tilde{f}(k,\tau') \ ,
\ee
with the assumption that the solution at the initial time $\tau_i$ is zero.  Up to this point no specific equation of state has been assumed and the results are completely general (within the context of the hydrodynamic model).

In general, the solution to the above equations cannot be written down in closed form.  As a consequence, the evolution operator generally would need to be computed numerically once an equation of state is specified.  The exception is when the drift matrix is constant in time.  That was the situation studied in \cite{Kapusta2012}, and it would be the situation here too if the equation of state is given by (\ref{TsquareEOS}) with $C=0$.  If the drift matrix was time-independent, ${\mathbf D} (k,\tau'') = {\mathbf D} (k)$,  then the solution would be 
\be 
\tilde{{\mb U}} (k; \tau, \tau')  = \left( \frac{\tau'}{\tau} \right)^{\mb{D} (k)} \ .
\label{Dconstant} 
\ee 
We will approximate $\tilde{{\mb U}}$ by the above expression with ${\mathbf D}$ evaluated at $\tau$.  This should be a very good approximation at large $T$ and/or $\mu$, and at least semi-quantitatively valid in any case.  (The reason for evaluating the drift matrix at $\tau$ instead of $\tau'$ is discussed more later.)  To avoid this approximation would require taking $C=0$, and it is arguable whether it would be a better approximation to the real physics.

The calculation is further simplified as the drift matrix is diagonalizable. The eigenvalues are:
\ba
d_0 &=&0 \ , \\
d_{\pm} &=& \alpha \pm \beta \ ,
\ea
with
\be \alpha \equiv \thalf\left(1-v_\sigma^2\right), \quad \beta^2 \equiv \alpha^2-v^2_\sigma k^2 \ . \ee
The evolution operator reads:
\be
 \tilde{{\mb U}} (k; \tau, \tau') =  \tilde{{\mb U}}_0 (k; \tau')\\
+ \frac{\left( \tau' /\tau \right)^{d_-}}{d_+ - d_-} \tilde{{\mb U}}_- (k; \tau')
 -   \frac{\left( \tau'/\tau \right)^{d_+}}{d_+-d_-} \tilde{{\mb U}}_+ (k; \tau') \ ,
\ee
where
\be
\tilde{{\mb U}}_0 (k; \tau') = \frac{1}{wv_\sigma^2 } 
\left(  
\begin{array}{ccc}
 n \mu v_s^2 & - \mu s v_s^2 & 0 \\  
- T n v_n^2 &  T s v_n^2 & 0 \\  
 0 & 0 & 0 
\end{array}
\right) \ ,
\ee
\be
\tilde{{\mb U}}_{\pm} (k; \tau') = \left( 
\begin{array}{ccc}
\displaystyle{\frac{v_n^2}{v_\sigma^2}\frac{Ts}{w}} d_{\mp} &  \displaystyle{\frac{v_s^2}{v_\sigma^2}\frac{\mu s}{w}} d_{\mp}
 & - ik \\[7pt]
\displaystyle{\frac{v_n^2}{v_\sigma^2}\frac{Tn}{w}} d_{\mp} & \displaystyle{\frac{v_s^2}{v_\sigma^2 }\frac{\mu n}{w}} d_{\mp}
 & - ik \displaystyle{\frac{n}{s} }\\[7pt]
   -ik v_n^2 \displaystyle{\frac{Ts}{w}}  & - ik v_s^2 \displaystyle{\frac{\mu s}{w}} 
 & - d_{\pm} 
\end{array}
\right) \ .
\ee
The solution to the inhomogeneous Langevin equation gives the response functions $\tilde{G}_X (k,\tau,\tau')$. They are calculated from
\be 
\left(
\begin{array}{c}
 \tilde{G}_s (k; \tau,\tau') \\
 \tilde{G}_n (k; \tau,\tau') \\
 \tilde{G}_\omega (k; \tau,\tau')
\end{array}
\right) = \tilde{{ \mb U}} (k; \tau,\tau') {\mb n}(k,\tau') 
\ee
to be
\begin{eqnarray}
\nonumber   \tilde{G}_s (k; \tau,\tau') & = &  -\frac{ik}{v_{\sigma}^2} \frac{\mu}{T} \left\{v_s^2 +  \left(v_{\sigma}^2 -v^2_s\right) \left(
 \frac{\tau'}{\tau} \right)^{\alpha} \left[ \frac{\alpha}{\beta} \sinh \left( \beta \ln \frac{\tau}{\tau'} \right) + \cosh \left( \beta \ln \frac{\tau}{\tau'}\right) \right]   \right\} \ , \\
\nonumber \tilde{G}_n (k; \tau,\tau') & = & \frac{ik}{v_{\sigma}^2} \left\{ v_n^2 +  \left(v_{\sigma}^2-v_n^2\right) \left(
 \frac{\tau'}{\tau} \right)^{\alpha} \left[ \frac{\alpha}{\beta} \sinh \left( \beta \ln \frac{\tau}{\tau'} \right) + \cosh \left( \beta \ln \frac{\tau}{\tau'}\right) \right]   \right\} \ ,\\
 \tilde{G}_\omega (k; \tau,\tau') & = & \frac{k^2}{\beta} \frac{s}{n} \left(v_{\sigma}^2-v_n^2\right) \left( \frac{\tau'}{\tau}\right)^\alpha \sinh \left( \beta \ln \frac{\tau}{\tau'} \right) \ .
\label{Gsolutions}
\end{eqnarray}
where, unless otherwise indicated, the variables on the right hand side are functions of $\tau$.  

Although $\delta n/s$ and $\delta s/s$ were the natural dimensionless variables to use when solving the equations of motion, it is very useful to look at the variables
\be
\frac{\delta P}{Ts} = v_n^2 \left(\frac{\delta s}{s}\right) + v_s^2 \frac{\mu}{T} \left(\frac{\delta n}{s}\right)
\ee
and
\be
\delta \left( \frac{1}{\sigma} \right) = \delta \left(\frac{n}{s}\right) = \left(\frac{\delta n}{s}\right) - \frac{n}{s} \left(\frac{\delta s}{s}\right) \ .
\ee
The corresponding response functions are
\be
\tilde{G}_P (k; \tau,\tau')  = ik \, \frac{\mu}{T}  \left(v_s^2-v_n^2\right) \left( \frac{\tau'}{\tau} \right)^{\alpha} 
\left[ \frac{\alpha}{\beta} \sinh \left( \beta \ln \frac{\tau}{\tau'} \right) + \cosh \left( \beta \ln \frac{\tau}{\tau'}\right) \right]
\ee
and
\be
\tilde{G}_{\sigma} = ik \, \frac{w}{Ts} \ .
\ee
The reason that these are interesting and relevant is that a change in pressure at constant entropy per baryon corresponds to a sound wave, while a change in entropy per baryon at constant pressure corresponds to diffusive heat flow, and so these are physically orthogonal variables.  This should be apparent mathematically as well.  Notice that $\tilde{G}_P$ contains only $\cosh$ and $\sinh$ terms; in the large $k$ limit these become sinusoidal oscillations.  On the other hand $\tilde{G}_{\sigma}$ has no such terms; in coordinate space it is the derivative of a Dirac delta function.  

Approximating the matrix $\mathbf{D}$ as a constant, as in Eq.~(\ref{Dconstant}), can be examined in light of its diagonalized form.  The approximation will be good if 
$| \partial \ln v_{\sigma}^2 / \partial \ln \tau | \ll 1$.  This is a good approximation because it is only the presence of a $T^2$ term in the pressure that causes $v_{\sigma}^2$ to deviate from its asymptotic value of 1/3.  For a static, uniform system no such approximation is necessary.  See Appendix~\ref{app:perturbations}.

In the presence of the fluctuating forces, the solution to the equations of motion for $X=\delta s/s, \delta n/s,\omega, \delta P/Ts, \delta \sigma$ reads:
\be 
\tilde{X} (k,\tau)= - \int_{\tau_i}^{\tau} \frac{d\tau'}{\tau'} \tilde{G}_X (k,\tau,\tau') \tilde{f} (k,\tau') \ .
\ee
The correlation function of the fluctuating force (\ref{bfluctuations}) is
\be 
\langle f(\tau_1, \xi_1) f(\tau_2, \xi_2) \rangle = \frac{2 \lambda(\tau_1)}{A \tau_1} \left[ \frac{n (\tau_1) T(\tau_1)}{s(\tau_1) w(\tau_1)} \right]^2 \delta(\tau_1 -\tau_2) \delta(\xi_1 - \xi_2) \ , 
\ee
with
\be 
\delta (x_1-x_2) = \frac{1}{A\tau_1} \delta(\tau_1-\tau_2) \delta(\xi_1-\xi_2) 
\ee
and where $A$ is the transverse area in the Bjorken model.  The Fourier transform is
\be 
\langle \tilde{f}(k_1,\tau_1) \tilde{f}(k_2,\tau_2)\rangle = \frac{4 \pi \lambda(\tau_1)}{A \tau_1} \left[ \frac{n (\tau_1) T(\tau_1)}{s(\tau_1) w(\tau_1)} \right]^2 \delta(\tau_1-\tau_2) \delta(k_1+k_2) \ . 
\ee
In Fourier space the correlation function of the fluctuating variables then reads
\ba
\langle \tilde{X} (k_1,\tau_1) \tilde{Y} (k_2,\tau_2) \rangle &=& \frac{4 \pi}{A} \delta(k_1+k_2) \int_{\tau_i}^{{\rm min}(\tau_1,\tau_2)} 
\frac{d\tau}{\tau^3} \lambda(\tau) \left[ \frac{n (\tau) T(\tau)}{s(\tau) w(\tau)} \right]^2 \nonumber \\
&\times& \tilde{G}_X (k_1;\tau_1,\tau) \tilde{G}_Y (k_2;\tau_2,\tau) \ . 
\ea
The equal-time correlator at the final or freeze out time $\tau_f$ is
\be 
\tilde{C}_{XY} (k; \tau_f) =\frac{4 \pi}{A} \int_{\tau_i}^{\tau_f}  \frac{d\tau}{\tau^3} \lambda(\tau) \left[ \frac{n (\tau) T(\tau)}{s(\tau) w(\tau)} \right]^2 
\tilde{G}_{XY} (k;\tau_f,\tau) \ , 
\ee
where
\be 
\tilde{G}_{XY} (k; \tau_f,\tau) = \tilde{G}_X (k; \tau_f,\tau) \tilde{G}_Y (-k;\tau_f,\tau) \ . 
\ee
Finally, the correlation function in space-time rapidity is
\be C_{XY} (\xi_1-\xi_2; \tau_f) = \frac{2}{A} \int_{\tau_i}^{\tau_f} \frac{d\tau}{\tau^3} \lambda(\tau)  \left[ \frac{n (\tau) T(\tau)}{s(\tau) w(\tau)} \right]^2  G_{XY} (\xi_1-\xi_2; \tau_f ,\tau) \ , \ee
where $G_{XY} (\xi_1-\xi_2; \tau_f ,\tau)$ is the inverse Fourier transform of $\tilde{G}_{XY} (k; \tau_f,\tau)$.  As mentioned earlier, we will ignore the background thermal conductivity $\lambda^b$ in our numerical results.  The final correlations due to thermal conduction would only be greater if $\lambda^b$ was included.

\begin{figure}[top]
\includegraphics[width=0.5\linewidth]{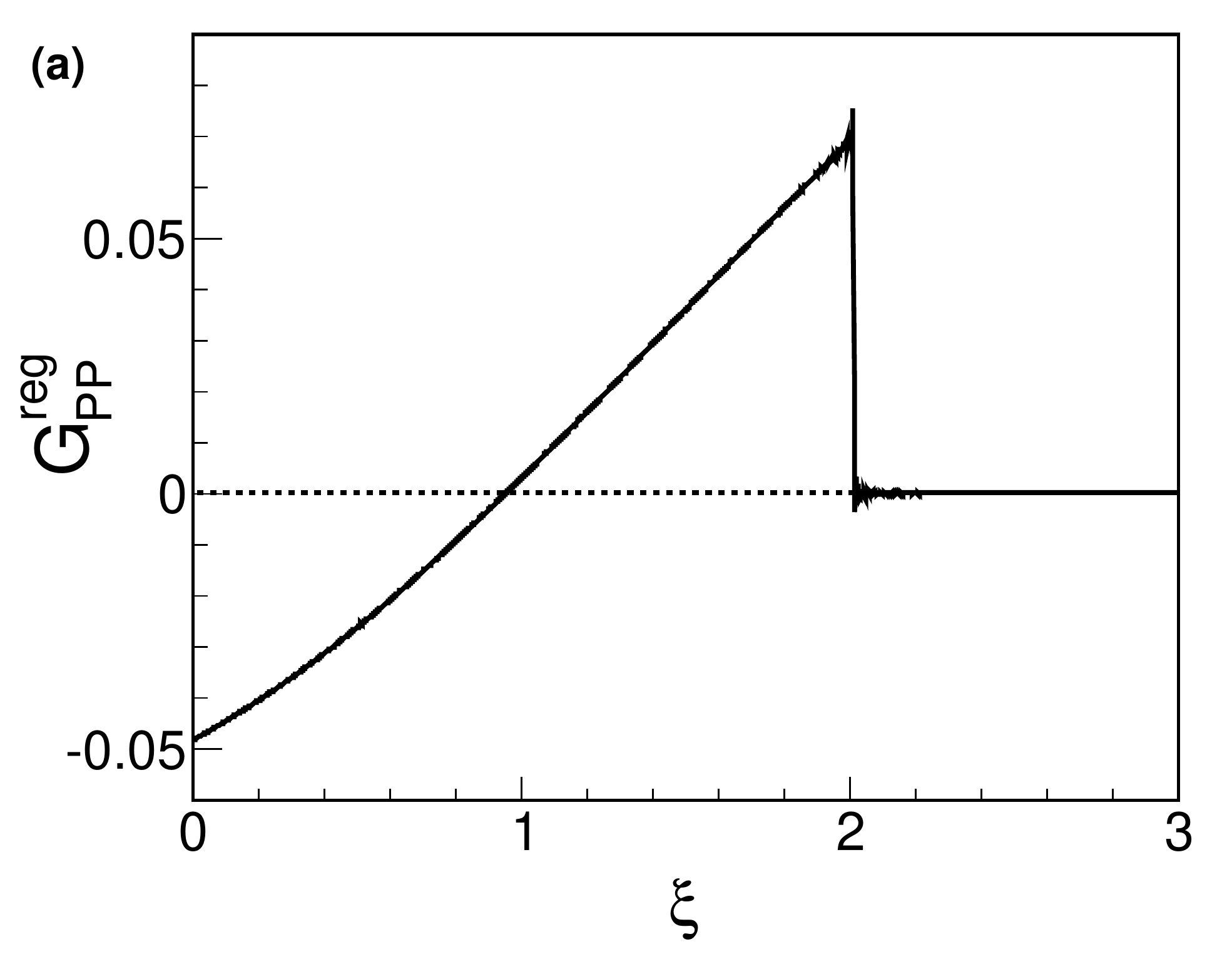}
\includegraphics[width=0.5\linewidth]{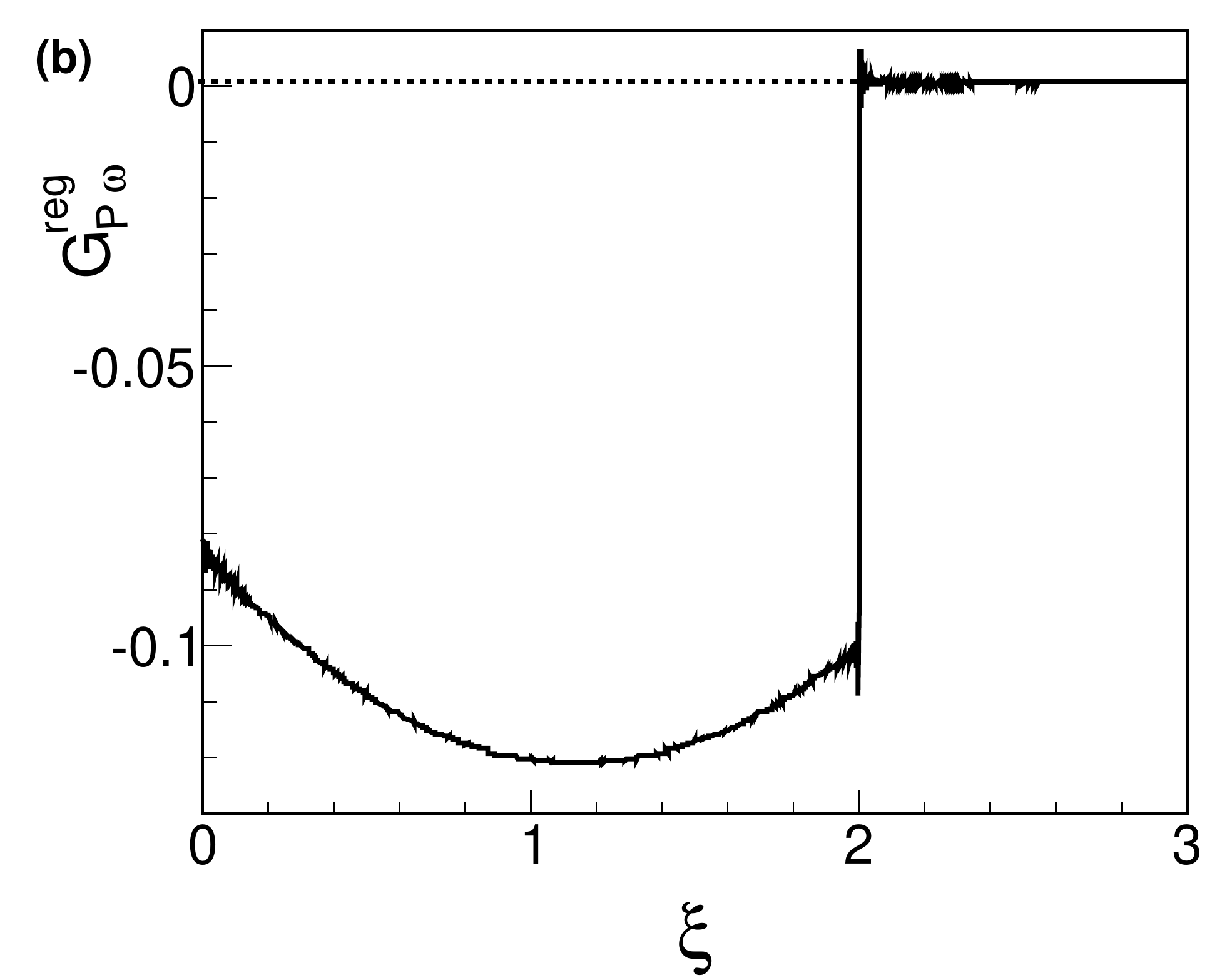}
\includegraphics[width=0.5\linewidth]{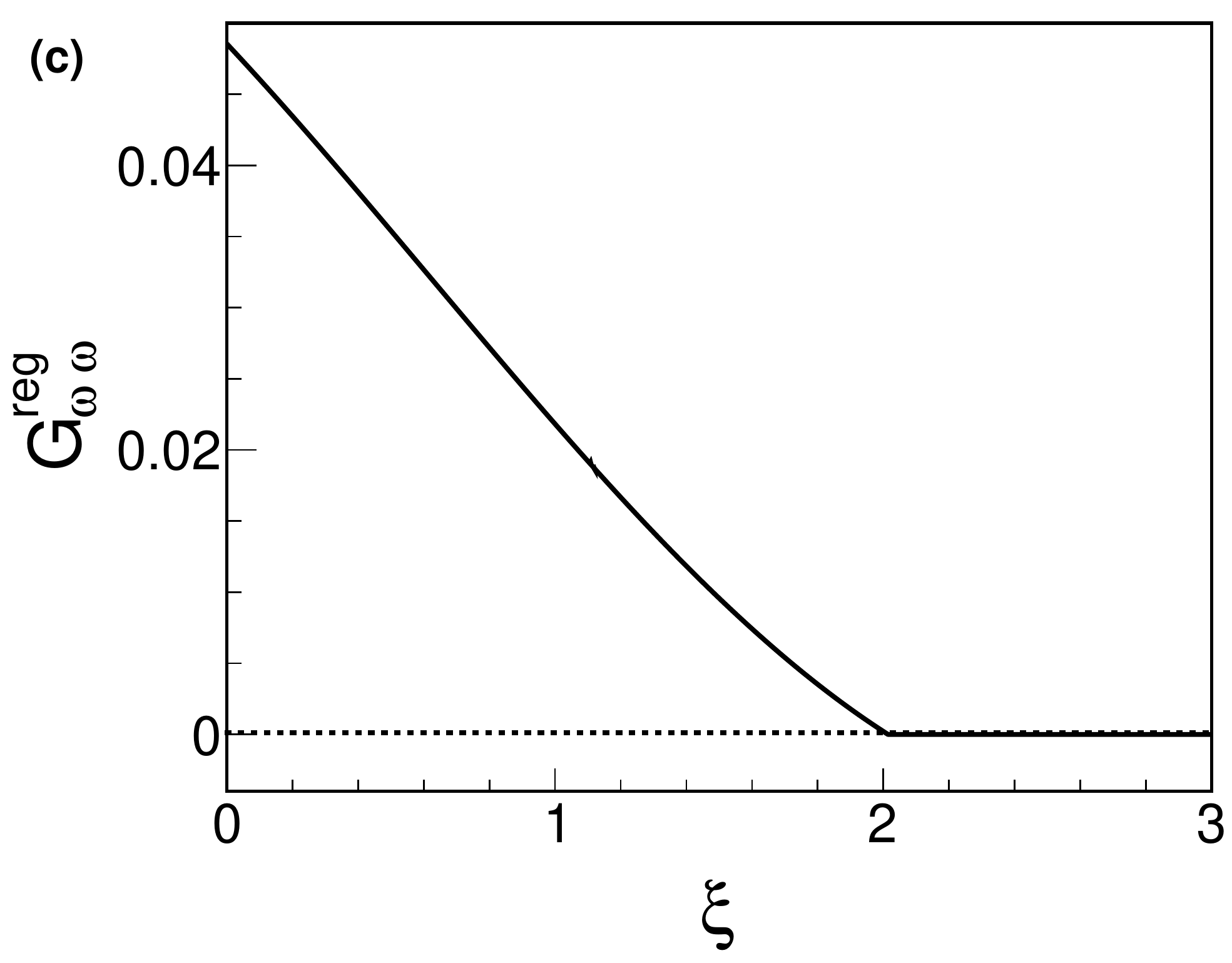}
  \caption{Regular part of the response functions: (a) $G^{reg}_{PP}$, (b) $G^{reg}_{P\omega}$ and (c) $G^{reg}_{\omega \omega}$. They represent a wake behind fronts that move at the speed of sound.}
  \label{fig:responsefunc}
\end{figure}

As discussed in some detail in \cite{Kapusta2012}, the response functions generally contain step functions, Dirac delta functions, and first and second derivatives of Dirac delta functions.  These are due to the fact that signals created by the fluctuating sources propagate at the speed of sound; the singular parts are located at the sound horizon in space-time rapidity.  The distance a signal can propagate in space-time rapidity between time $\tau_1$ and $\tau_2$ is
\be
\Delta \xi=\int_{\tau_1}^{\tau_2} \frac{d\tau}{\tau} v_{\sigma}(\tau) \ .
\label{xidistance}
\ee
Two points separated by twice this distance would receive a signal at the same time if it was sent from a source midway between them.  As in \cite{Kapusta2012}, one can subtract the singular terms to study the regular behavior of this function.
\be 
\tilde{G}^{{\rm reg}}_{XY} (k; \tau_f,\tau) = \tilde{G}_{XY} (k; \tau_f,\tau) - \tilde{G}^{{\rm sing}}_{XY} (k; \tau_f,\tau) \ . 
\ee
We show some examples of ${G}^{{\rm reg}}_{XY} (\xi; \tau_f,\tau_i)$ in Fig. \ref{fig:responsefunc}.  These represent the wake behind the propagating sound front.  The singular expansions of these coefficients are given in Appendix~\ref{app:singexpansion}. Note that the subtraction of the singular part is only made for illustration purposes. The final particle correlation contains the entire $\tilde{G}_{XY} (k; \tau_f,\tau)$ function.  It is noteworthy that if we approximate the drift matrix at the time $\tau'$ instead of at $\tau$ there would be a small singular piece in $\tilde{G}_{PP}$ at the distance given by (\ref{xidistance}) in addition to at twice that distance.  This is unphysical, and reaffirms the choice of the time $\tau$ rather than $\tau'$.   

It is interesting to observe that there are no sound waves emitted if $v_{\sigma}=v_s=v_n$, only pure diffusion, as may be seen from Eq.~(\ref{Gsolutions}).  Such is the case for the simple equation of state $P=v^2\epsilon - B$ where $v^2$ is a constant.  Sound waves are generated only if the pressure responds differently to variations in energy density depending on whether the entropy density, the baryon density, or the entropy per baryon is held fixed.  It is immediately apparent that QCD does not have the property that all three speeds are equal.  The equation of state used for the space-time evolution in this article, given by Eq. (\ref{TsquareEOS}), has different speeds because it contains a $T^2$ term in the pressure.  For very large energy densities it has a decreasing influence, and all three speeds approach $1/\sqrt{3}$; see Appendix~\ref{app:speedsound}.

\section{Phenomenology} 
\label{sec:pheno}

In this section we consider observable consequences of fluctuations caused by the existence of a critical point in the QCD phase diagram.  Although direct comparison to experiment is not to be expected, we will find that the effects are quite significant and definitely worthy of further study.

The analysis follows that described in \cite{Kapusta2012} very closely. For this reason, we only give the key steps here.  The number of particles with degeneracy $d$ per phase-space volume is
\be 
\frac{dN}{d^3x d^3p} = \frac{d}{(2\pi)^3} f(\mb{x},\mb{p}) \ , 
\ee
with
\be 
f(\mb{x},\mb{p}) = {\rm e}^{-(u \cdot p-\mu)/T} 
\ee
being the Boltzmann distribution function. The four-velocity of the fluid cell is
\be 
u^{\mu} =\left(\cosh(\xi+\omega),\mb{u}_{\perp},\sinh(\xi+\omega)\right)  \ . 
\ee
The distribution of particles at the freeze-out surface $\Sigma_f$ is given by the Cooper-Frye formula
\be
E \frac{dN}{d^3p} = d \int_{\Sigma_f} \frac{d^3\sigma_{\mu}}{(2\pi)^3} \ p^{\mu} f(\mb{x},\mb{p}) \ . 
\ee
In the hydrodynamic model being used here
\be 
d^3\sigma_\mu \ p^{\mu} = \tau_f \ d\xi \ d^2x_{\perp} m_\perp \cosh(y-\xi) \ . 
\ee
The variable $y$ represents the particle rapidity
\be 
p^\mu = (m_\perp \cosh y, \mb{p}_\perp, m_\perp \sinh y) \ , 
\ee
where
\be 
m_\perp = \sqrt{m^2+p_\perp^2} 
\ee
is the transverse mass.  The number of particles per unit rapidity is then
\be 
\label{eq:distro} \frac{dN}{dy} =\frac{d A \tau_f }{(2\pi)^3} \ \int d\xi \cosh(y-\xi) \int d^2p_\perp m_\perp 
\exp\left\{-[(\cosh(y-\xi-\omega) m_\perp-\mu)/T]\right\} \ , 
\ee
where the integration over $\mb{x}_{\perp}$ gives the transverse area of the collision $A$.
If we neglect all fluctuations ($\delta T=\delta \mu=\omega=0$) we get the average of $dN/dy$ as
\be 
\Big\langle \frac{dN}{dy} \Big\rangle = \frac{d A \tau_f}{(2 \pi)^2} {\rm e}^{\mu/T} \int_{-\infty}^\infty d\xi \ \cosh(y-\xi) \int dp_\perp p_\perp m_\perp 
\exp\left\{-[\cosh(y-\xi) m_{\perp}]/T\right\}  \ .
\ee
In order to perform the integration over $p_\perp$ we use the following formulas:
\ba
\label{eq:integration1} \int dp_\perp p_\perp m_{\perp} {\rm e}^{-c m_{\perp}} &= &  \frac{1}{c^3} \ {\rm e}^{-cm} [2+2cm+(cm)^2] \equiv \frac{1}{c^3} \Gamma(3,cm) \ , \\
\label{eq:integration2} \int dp_\perp p_\perp m_{\perp}^2 {\rm e}^{-c m_{\perp}} &= & \frac{1}{c^4} \ {\rm e}^{-cm} [6+6cm+3(cm)^2+(cm)^3] \equiv \frac{1}{c^4} \Gamma(4,cm) \ .
\ea
At the freeze-out time we obtain
\be 
\frac{dN}{dy} =\frac{d A \tau_f T_f^3}{4 \pi^2} {\rm e}^{\mu_f/T_f} \int_{-\infty}^{\infty} \frac{dx}{ \cosh^2 x} 
\Gamma \left(3, \frac{m}{T_f} \cosh x \right) \ .
\ee
Now we consider fluctuations of $dN/dy$ and eventually its two-point correlation. To do so, we expand the exponential term
in (\ref{eq:distro}) to first order in fluctuations around the freeze out value:
\ba 
T & =&T_f +\delta T (\tau_f,\xi) \ , \\
\mu & = & \mu_f +\delta \mu (\tau_f,\xi) \ .
\ea
The Boltzmann factor becomes
\bd
\exp\left\{-[(\cosh(y-\xi-\omega) m_\perp-\mu)/T]\right\} \rightarrow  \exp\left\{-[(\cosh(y-\xi) m_\perp-\mu_f)/T_f]\right\}
\ed
\be
\times \left\{ 1 + \frac{\delta T (\xi)}{T_f^2} \left[ m_\perp \cosh(y-\xi)  - \mu_f \right] 
+ \omega (\xi) \frac{m_\perp}{T_f} \sinh(y-\xi) +  \frac{\delta \mu (\xi)}{T_f} \right\} \ ,
\ee
where the fluctuations are understood to be evaluated at $\tau_f$.  The fluctuation in the number of particles per unit rapidity is then
\bd
\delta \left( \frac{dN}{dy} \right) = \frac{dA}{(2\pi)^3} \tau_f \int d\xi \cosh(y-\xi) \int d^2p_\perp m_\perp 
\exp\left\{-[(\cosh(y-\xi) m_\perp-\mu_f)/T_f]\right\}
\ed
\be
\times \left\{\frac{\delta T (\xi)}{T_f^2} \left[ m_\perp \cosh(y-\xi)  - \mu_f \right] 
+ \omega (\xi) \frac{m_\perp}{T_f} \sinh(y-\xi) +  \frac{\delta \mu (\xi)}{T_f} \right\}\ .
\label{eq:fluc} 
\ee
We need to replace the fluctuations $\delta T$ and $\delta \mu$ by $\delta s$ and $\delta n$. We use
\ba 
\delta T &=& \frac{\chi_{\mu \mu}}{\Delta} \ \delta s - \frac{\chi_{T \mu}}{\Delta} \ \delta n  \ , \nonumber \\
\delta \mu & = & - \frac{\chi_{T \mu}}{\Delta} \ \delta s + \frac{\chi_{T T}}{\Delta} \ \delta n \ , 
\ea
where $\chi_{TT} = \partial^2P(T,\mu)/\partial T^2$,  $\chi_{T\mu} = \partial^2P(T,\mu)/\partial T\partial \mu$,  and $\chi_{\mu\mu} = \partial^2P(T,\mu)/\partial \mu^2$, and where
\be 
\Delta = \chi_{TT} \chi_{\mu \mu} - \chi_{T \mu}^2 \ . 
\ee
We now perform the integration over $\mb{p}_{\perp}$ with the help of (\ref{eq:integration1}) and (\ref{eq:integration2}). The fluctuation of $dN/dy$ reads:
\be \delta \left( \frac{dN}{dy} \right) = \frac{dA\tau_f T_f^2}{4\pi^2}  \ {\rm e}^{\mu_f/T_f}
 \int d\xi \ \left[  \frac{\delta s}{s} F_s (y-\xi)  + \omega  F_\omega(y-\xi) + \frac{\delta n}{s} F_n (y-\xi) \right] \ .
\ee
Here we have introduced the functions
\begin{eqnarray}
F_s (x) & \equiv &  \frac{s \chi_{\mu \mu}}{\Delta \cosh^2 x} \Gamma\left(4,\frac{m}{T_f} \cosh x\right) -  \frac{s\chi_{T\mu}+s\chi_{\mu\mu}\mu_f/T_f}{\Delta \cosh^2 x} 
\Gamma\left(3,\frac{m}{T_f} \cosh x \right) \ , \\ 
 F_\omega(x) & \equiv & \frac{T_f \tanh x}{\cosh^2 x} \Gamma\left(4,\frac{m}{T_f} \cosh x \right) \ , \\
 F_n (x)& \equiv & -\frac{s \chi_{T \mu}}{\Delta \cosh^2 x} \Gamma \left(4,\frac{m}{T_f} \cosh x \right) + \frac{s\chi_{TT}+s\chi_{T\mu} \mu_f/T_f}{\Delta \cosh^2 x} \Gamma \left(3,\frac{m}{T_f} \cosh x \right) \ . 
\end{eqnarray}
Finally, we construct the rapidity correlator:
\bd \Big\langle \delta \left( \frac{dN}{dy_1} \right) \delta \left( \frac{dN}{dy_2} \right) \Big\rangle = \left( \frac{dA\tau_f T_f^2}{4\pi^2} \right)^2
 {\rm e}^{2 \mu_f/T_f} \int d\xi_1  \int d\xi_2
\ed
\be
\times \sum_{X,Y} F_X(y_1-\xi_1) F_Y(y_2-\xi_2) C_{XY}(\xi_1-\xi_2;\tau_f) \ ,
\ee
where the sum runs over $\delta s/s$, $\delta n/s$, and $\omega$, and where
\be 
C_{XY} (\xi_1-\xi_2; \tau_f) = \langle X(\xi_1;\tau_f) Y(\xi_2;\tau_f) \rangle \ . 
\ee

To simplify calculation of the 2-particle correlation function it is convenient to use the Fourier transforms.
\be
\Big\langle \delta \left( \frac{dN}{dy_1} \right) \delta \left( \frac{dN}{dy_2} \right) \Big\rangle = \left( \frac{dA\tau_f T_f^2}{4\pi^2} \right)^2
{\rm e}^{2 \mu_f/T_f} \int \frac{dk}{2\pi} {\rm e}^{ik\Delta y} \sum_{X,Y} 
\tilde{F}_X(k) \tilde{F}_Y(-k) \tilde{C}_{XY}(k;\tau_f) \ , 
\ee
with $\Delta y=y_1-y_2$.  The expression for $\tilde{C}_{XY}$ is:
\be 
\tilde{C}_{XY} (k) = \frac{4\pi}{A} \int_{\tau_i}^{\tau_f} \frac{d\tau}{\tau^3} \lambda(\tau) \left( \frac{nT}{sw} \right)^2 \tilde{G}_X (k;\tau_f,\tau)
\tilde{G}_Y (-k;\tau_f,\tau) \ . 
\ee
To eliminate the transverse area we normalize the correlation function by the event average of $dN/dy$ to obtain
\be 
\Big\langle \delta \left( \frac{dN}{dy_1} \right) \delta \left( \frac{dN}{dy_2} \right) \Big\rangle \Big\langle \frac{dN}{dy} \Big\rangle^{-1} = \frac{d \tau_f T_f^2}{2\pi^2}
{\rm e}^{\mu_f/T_f} \frac{C(\Delta y)}{N \left(m/T_f \right)} \ , 
\ee
where 
\be
C(\Delta y)= \int dk \ {\rm e}^{ik\Delta y} \sum_{X,Y} \tilde{F}_X(k) \tilde{F}_Y (-k) \int_{\tau_i}^{\tau_f} \frac{d\tau}{\tau^3}
\lambda \left( \frac{nT}{sw} \right)^2 \tilde{G}_X (k;\tau_f,\tau) \tilde{G}_Y (-k;\tau_f,\tau)  
\ee
and
\be 
N \left(m/T_f \right) \equiv \int_{-\infty}^{\infty} \frac{dx}{\cosh^2 x} \Gamma \left( 3,\frac{m}{T_f} \cosh x\right) \ . 
\ee

\begin{figure}
\includegraphics[width=0.7\linewidth]{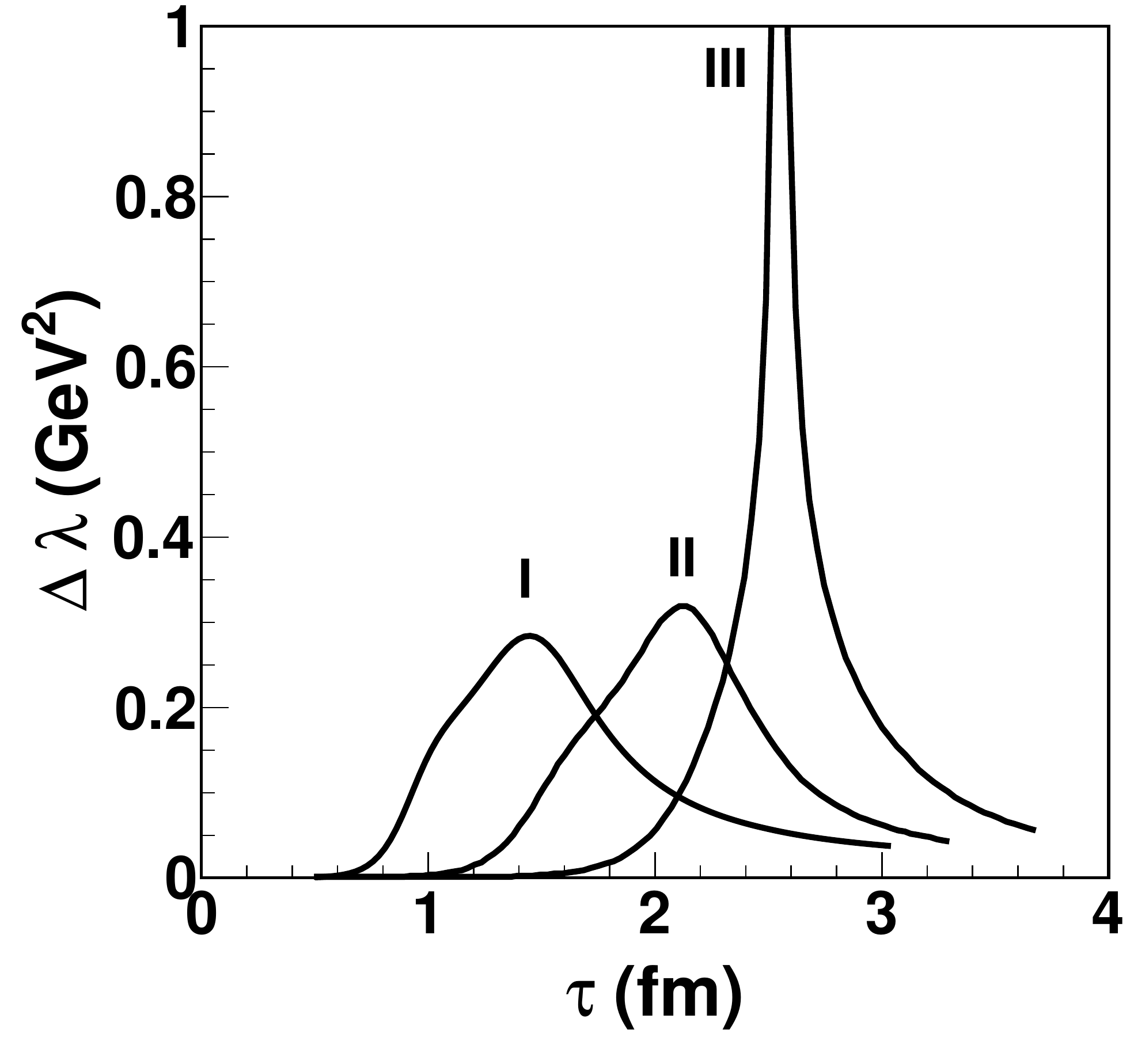}
  \caption{Thermal conductivity as a function of proper time and trajectory.}
  \label{fig:thermalcond}
\end{figure}

\begin{figure}
\includegraphics[width=0.7\linewidth]{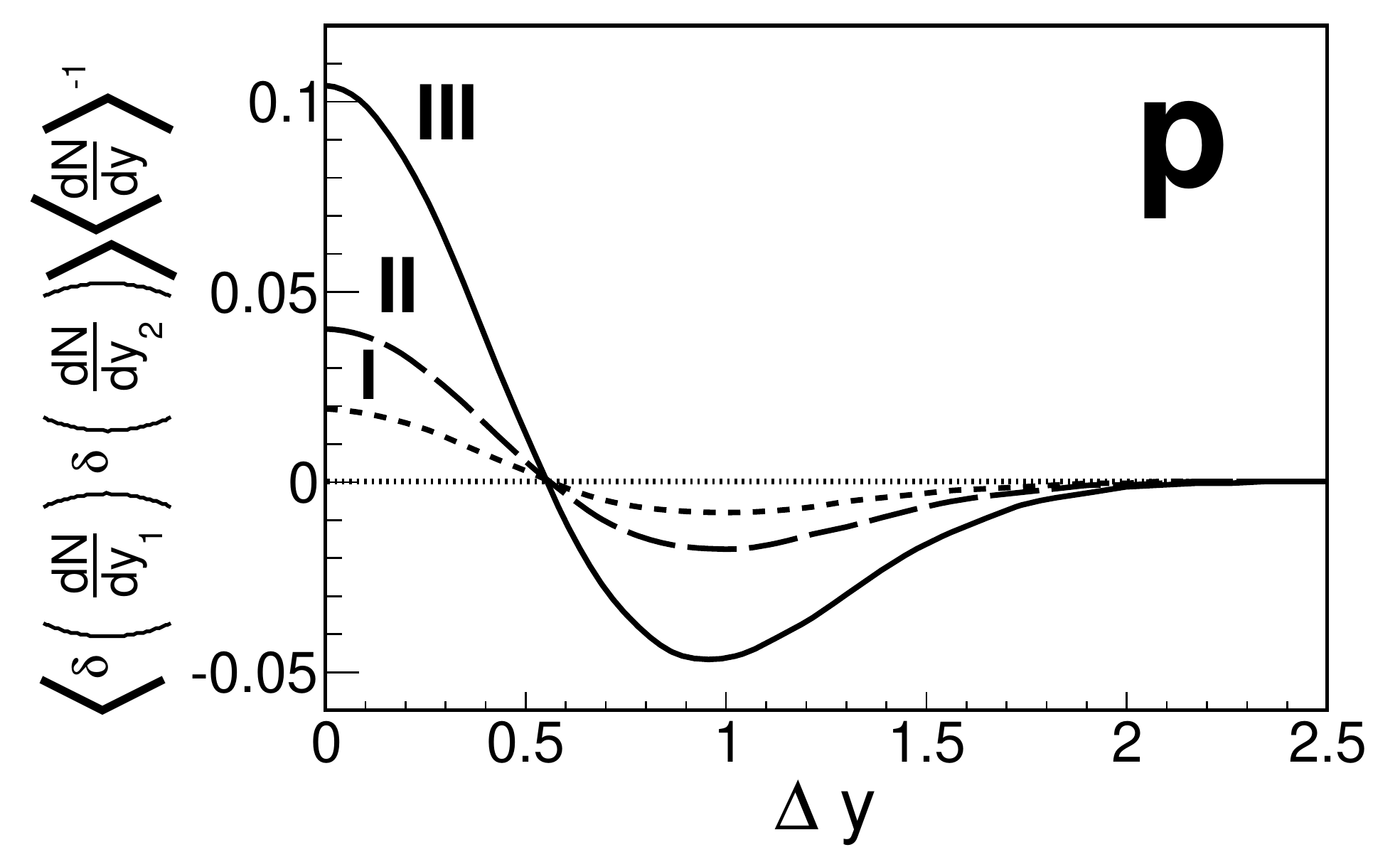}
  \caption{Particle correlation function for protons.}
  \label{fig:protoncorrelation}
\end{figure}

\begin{figure}
\includegraphics[width=0.7\linewidth]{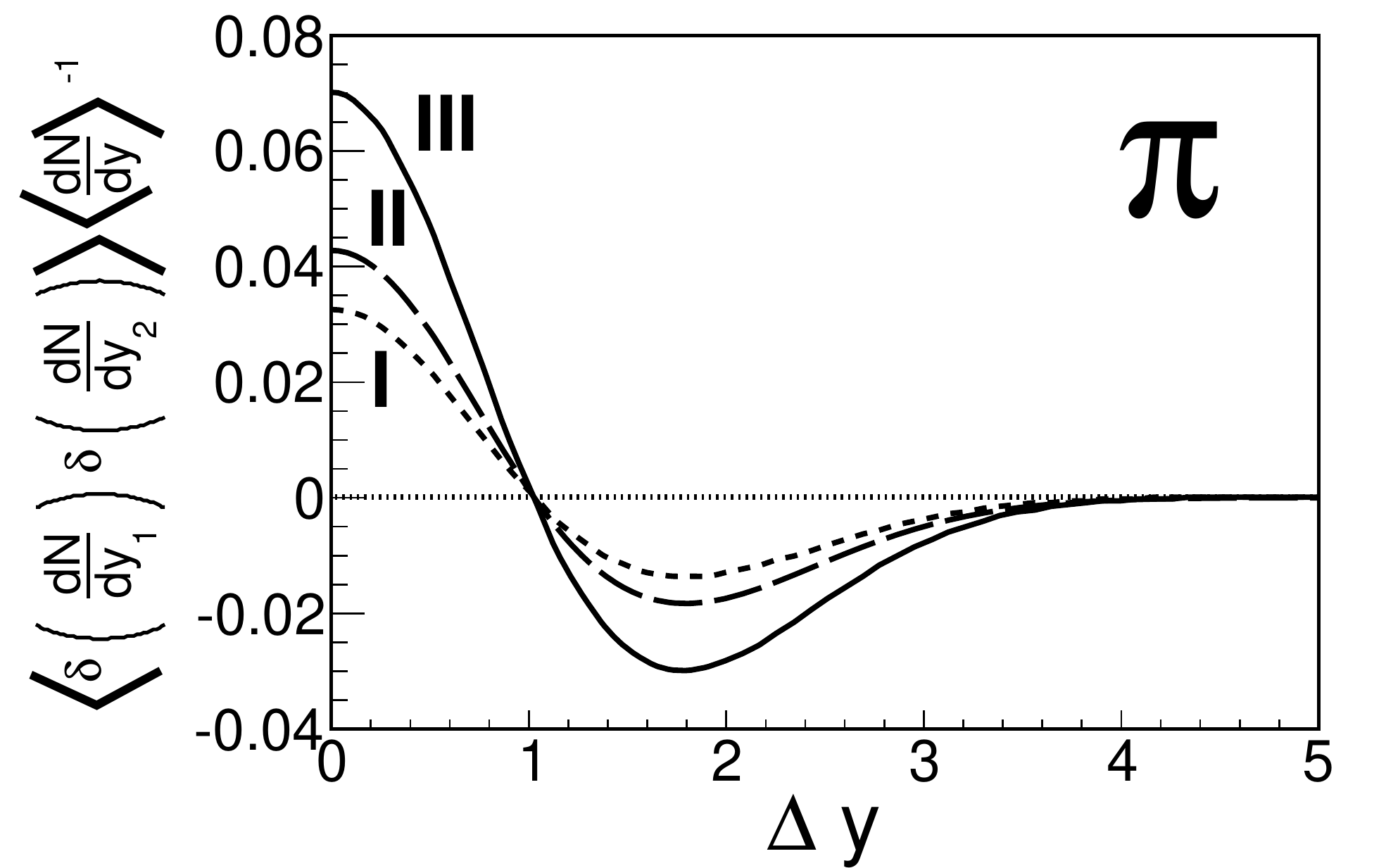}
\includegraphics[width=0.7\linewidth]{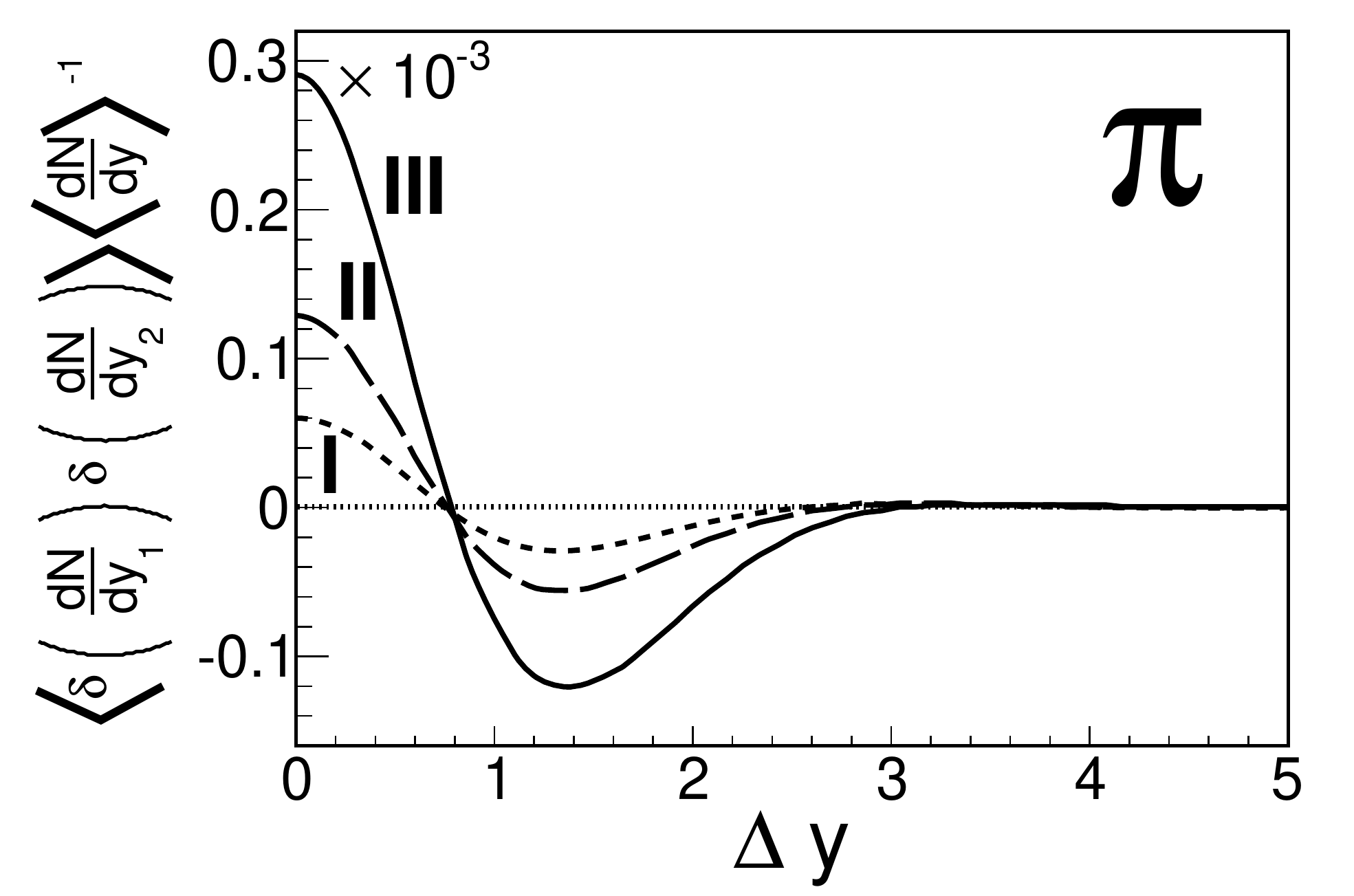}
  \caption{Particle correlation function for pions at zero chemical potential. (a) Non-zero fluctuations of chemical potential. (b) No fluctuations of chemical potential.}
  \label{fig:pioncorrelation}
\end{figure}

Now we are prepared to study the influence of the hydrodynamic fluctuations on the 2-particle correlation functions.  Figure \ref{fig:thermalcond} shows the thermal conductivity as a function of time for the three trajectories chosen earlier.  As the trajectory passes closer to the critical point the enhanced thermal conductivity is probed ever more closely. 
The effect of this on the 2-particle correlation functions is shown in Fig. \ref{fig:protoncorrelation} for protons ($d=2$, $m_p=939$ MeV). For charged pions ($d=2$, $m_{\pi}=138$ MeV, zero chemical potential) we show in Fig. \ref{fig:pioncorrelation} the 2-particle correlation function with no chemical potential fluctuation ($\delta \mu=0$) (left panels) and with pion chemical potential fluctuation equal to that of the protons (right panels). Which one is closer to reality cannot be determined without explicitly introducing a conserved electric current on the same footing as the baryon current.  The presence of additional fluctuations enhances the magnitude of the correlation between particles. Both pions and protons exhibit the influence of the thermal conductivity.  The shape is approximately the same for all trajectories.  The protons have a maximum at $\Delta y = 0$ and a minimum near $\Delta y = 0.95$; the pions have a maximum at $\Delta y = 0$ and a minimum near $\Delta y = 1.7$ (with chemical 
potential fluctuations) or near 1.3 (without chemical potential fluctuations).  The magnitude of the correlation, on the other hand, increases dramatically as one goes from trajectory I to II to III.  As mentioned earlier, the boost invariant hydrodynamical model is not sufficiently realistic to consider any comparison to experiment, especially since one needs a large initial baryon number.  Nevertheless, the size of the effect in this model bodes well for future theoretical studies and experiments.


\section{Summary and conclusions}
\label{sec:conclusion}

We applied mode coupling theory, together with a parameterization of the equation of state that incorporates the correct critical exponents and amplitudes, to develop a model for the thermal conductivity in the vicinity of the critical point.  This contribution to the thermal conductivity incorporates the correct critical behavior but can be used in the non-asymptotic region as well.  The thermal conductivity quantifies the strength of particular hydrodynamic fluctuations via the fluctuation-dissipation theorem. To obtain insight into what effects might result in heavy ion collisions as a consequence of the critical point, we studied a simple boost-invariant hydrodynamic model.  We conservatively assumed that the entropy per baryon created in a heavy ion collision is too large for an adiabat to enter the mixed phase.  The thermal conductivity along the adiabatic trajectory is enhanced the closer the trajectory comes to the critical point.  The flyby of the critical point results in fluctuations in the 
temperature, baryon chemical potential, and local flow velocity which evolve with time and are not the same as fluctuations in the initial conditions.  We found that the growth of the thermal conductivity near the critical point implies the existence of two-particle correlations over 2 units of rapidity for protons.  The strength of this correlation increases the closer the expansion trajectory comes to the critical point. In particular, the magnitude of the correlation is directly proportional to the time-integrated history of the thermal conductivity, making the correlation a sensitive probe of the thermal conductivity and of the presence of a critical point.  With the inclusion of other transport coefficients (shear and bulk viscosities) this correlation can be further enhanced. We have found that the fluctuations in the baryon density --in particular the chemical potential fluctuations-- are the most important effect for the magnitude of correlations compared to the thermal and velocity fluctuations 
alone. For this reason, we think that the usually neglected baryon diffusion coefficient --alternatively, the thermal conductivity-- could be of interest for accessing the critical behavior by two-particle correlations.  However, their practical use relies on the ability of the heavy-ion factories to produce trajectories that pass close to the critical point.

There are many natural extensions to this study.  For example, inclusion of the regular part of the thermal conductivity would increase the magnitude of the fluctuations and hence correlation functions in the final state observables.  To quantify fluctuations due to electric charge, pions for example, requires the introduction of the electric charge current in addition to the baryon current.  Introduction of the strangeness degree of freedom --the kaon multiplicity is typically larger than proton multiplicity-- would also enhance the correlation function as one includes additional fluctuations in the strange chemical potential.  Inclusion of the thermal conductivity into the fluid equations of motion, not only in the fluctuations, should certainly be done. This could potentially increase correlations because it couples some of the fluctuations among them in such a way that the drift matrix does not contain any nonvanishing element. The effect of this modification for the shear and bulk viscosities has been 
studied in \cite{Kapusta2012} leading to a smoothing of the singularities appearing at the sound horizon.

An extension to the region of first order phase transition would be interesting.  Decreasing the initial entropy per baryon below its value at the critical point would result in instabilities and phase separation, with probably greater consequences than a flyby of the critical point as studied here.  The importance of the thermal conductivity in this regard has already been noted \cite{Skokov2010}.  Ultimately a 3+1 dimensional calculation for the fireball evolution is necessary to study the hydrodynamical correlations not only in rapidity but also as a function of the azimuthal angle, analogous to what was done in Ref. \cite{Shuryak:2009cy}-\cite{Staig:2011wj} for initial state fluctuations. Their effects on the flow harmonics and the angular correlation of particles ought to be important. Such a study is of course much more difficult than the one presented here since one must perform extensive numerical calculations.

\acknowledgments

This work was supported by the U.S. DOE Grant No. DE-FG02-87ER40328 and grant FPA2011-27853-C02-01.  J. M. T.-R. was also supported by an FPU grant from the Spanish Ministry of Education. He thanks the University of Minnesota for kind hospitality during a 3 month visit in 2012.

\newpage

\appendix

\section{\label{app:baryoncurrent} Baryon Current in a Gradient Expansion}

In the Landau-Lifshitz definition of flow velocity the baryon current is modified by a term linear in a gradient and proportional to the thermal conductivity.
\be
J^{\mu} = nu^{\mu} + \lambda \left(\frac{nT}{w}\right)^2 \Delta^{\mu} (\beta \mu) \ ,
\ee
where $\beta = 1/T$ and $\Delta^{\mu} = \partial^{\mu}- u^{\mu}(u \cdot \partial)$.  The tensorial structure is determined by the requirement that the dissipative term be orthogonal to the flow velocity so that $u \cdot J = 0$.  Going to second order in a gradient expansion yields
\be
J^{\mu} = nu^{\mu} + \lambda \left(\frac{nT}{w}\right)^2 \Delta^{\mu} \left[ (1-\tau_B u \cdot \partial) \beta \mu\right] \ ,
\ee
where $\tau_B$ is a new time constant which is in principle dependent on $T$ and $\mu$.  For baryon diffusion at fixed temperature and in the local rest frame this results in a modified diffusion equation.
\be
\frac{\partial \delta \mu}{\partial t} =  D_B \nabla^2 \delta \mu - \tau_B D_B \nabla^2 \frac{\partial \delta \mu}{\partial t} \ .
\ee
This is a causal equation, unlike the conventional diffusion equation.  It should be mentioned that in the Bjorken hydrodynamic model that dissipative term in the baryon current vanishes identically because $\Delta^{\mu}$ is acting on a function of $\tau$ only.  The dissipative term would contribute to the linearized fluctuation equations, but the resulting fluctuations would be one higher order in $\lambda$.

\section{\label{app:speedsound} Speed of Sound}

The adiabatic speed of sound (constant $\sigma = s/n$) can be computed for (\ref{TsquareEOS}).  It is $v_{\sigma}^2 = \oneth - \Delta v_{\sigma}^2$ with
\be
\Delta v_{\sigma}^2 = \frac{2}{9} \frac{CT}{d(T,\mu)}
\frac{\sigma(A_2 T^2 + 6 A_0 \mu^2) -2A_2 \mu T}{\sigma T + \mu} \ ,
\ee
where
\be
d(T,\mu) = 2 A_4 A_2 T^4 + (12 A_4 A_0 - A_2^2) \mu^2 T^2 + 2 A_2 A_0 \mu^4 - \oneth C (A_2 T^2 + 6 A_0 \mu^2) \ .
\ee
Notice that $\Delta v_{\sigma}^2$ is proportional to $C$ and hence vanishes for a conformally invariant equation of state.  Since $ds = \sigma dn$, the speed at constant baryon density, $v_n^2 \equiv (\partial P/\partial \epsilon)_n$, can be obtained by taking $\sigma \rightarrow \infty$.  The speed at constant entropy density, $v_s^2 \equiv (\partial P/\partial \epsilon)_s$, can be obtained by taking $\sigma \rightarrow 0$.  It is easily verified that
\be
v_{\sigma}^2 = \frac{v_n^2 T s + v_s^2 \mu n}{w} \ ,
\ee
a relationship that is independent of the specific equation of state.  Of course waves do not physically propagate at constant $n$ or $s$, only at constant $\sigma$, but these definitions are useful for various intermediate steps in various applications.  For example, a thermodynamic identity is
\be
\left(\frac{\partial \mu}{\partial T}\right)_n = \frac{v_n^2 c_V - s}{n} \ ,
\ee
so that
\be
c_P - c_V = v_n^4 c_V^2 T \chi_B \ .
\ee

\section{\label{app:perturbations} Perturbations of a Static Uniform System}

It is instructive to be reminded of perturbations of a static, uniform system at rest.  The analog of the boost invariant hydrodynamics used as an example in the text is a three dimensional system with perturbations dependent on $z$ and $t$ but independent of $x$ and $y$.  The resulting equations for the fluctuations are solved straightforwardly since the drift matrix is time-independent.  The response functions correspond directly to those given in Section IV with the following replacements: $\alpha \rightarrow 0$, $\cosh \rightarrow \cos$, $\sinh \rightarrow \sin$, $\ln(\tau/\tau') \rightarrow t-t'$, and $\beta \rightarrow v_{\sigma}k$.  Also the function $\omega(\xi,\tau)$ is replaced by the velocity perturbation $v(z,t)$ in the $z$-direction.
\ba
\tilde{G}_s &=& -\frac{ik}{v_{\sigma}^2} \frac{\mu}{T} \left\{ v_s^2 + \left(v_{\sigma}^2 - v_s^2 \right) \cos[k v_{\sigma}(t-t')]\right\} \ , \nonumber \\
\tilde{G}_n &=& \frac{ik}{v_{\sigma}^2} \left\{ v_n^2 + \left(v_{\sigma}^2 - v_n^2 \right) \cos[k v_{\sigma}(t-t')]\right\} \ , \nonumber \\
\tilde{G}_v &=& \frac{k}{v_{\sigma}} \frac{s}{n} \left(v_{\sigma}^2 - v_n^2 \right) \sin[k v_{\sigma}(t-t')] \ , \nonumber \\
\tilde{G}_P &=& ik \frac{\mu}{T}  \left(v_s^2 - v_n^2 \right) \cos[k v_{\sigma}(t-t')] \ , \nonumber \\
\tilde{G}_{\sigma} &=& ik \frac{w}{Ts} \ .
\ea
From these it is immediately clear that a pressure disturbance will travel with the speed of sound $v_{\sigma}$ while disturbances in the entropy per baryon are local in coordinate space.

\section{\label{app:singexpansion} Singular Part of the Response Functions}

The response functions $\tilde{G}_{XY} (k;\tau,\tau')$ contain terms that produce singularities when the inverse Fourier transform is performed. The general expansion
of the singular part is obtained by a Laurent expansion in $1/k$ and retaining the regular terms at $k=0$. For the cases $XY=ss,nn,sn,\omega\omega$, the general
expansion of the singular part reads:
\begin{eqnarray}
  \nonumber \tilde{G}^{sing}_{XY} (k;\tau,\tau') & = & (a_1 k^2 + b_1) + (a_2 k^2 +b_2) \cos (2 v_{\sigma} L k) + \frac{a_3 k^2 + b_3}{k} \sin (2 v_{\sigma} L k) \\
\label{eq:singexp1} & + & (a_4 k^2 +b_4) \cos ( v_{\sigma} L k) + \frac{a_5 k^2 + b_5}{k} \sin ( v_{\sigma} L k) \ ,
\end{eqnarray}
with $L\equiv\ln(\tau/\tau')$. 
The coefficients $a_i$,$b_i$ are polynomials in $L$:
\be a_i= \sum_j a_{ij} L^j \ , \quad b_i= \sum_j b_{ij} L^j \ , \ee
whose coefficients $a_{ij}$ and $b_{ij}$ depend on some thermodynamical quantities. The expansions of $a_i$ and $b_i$ are represented 
in Tables~\ref{tab:Gsn} and \ref{tab:Goo} for the cases $XY=sn$ and $XY=\omega \omega$. Note that for the case $XY=ss$ and $XY=nn$ one
should use Table~\ref{tab:Gsn} and make $D \rightarrow A, F \rightarrow C, E \rightarrow B$ and $A \rightarrow D, C \rightarrow F, B \rightarrow E$, respectively.
\begin{table}[h]
\begin{center}
\begin{tabular}{||c|c|c|c|c||}
\hline \hline
 & $j=0$ & $j=1$ & $j=2$ & $j=3$ \\
\hline
$a_{1j}$ & $-AD - CF/2$& $0$&$0$ &$0$ \\
$b_{1j}$ & $- BE/2v_{\sigma}^2$ & $0$ & $0$ & $0$ \\
$a_{2j}$ & $-CF/2$& $0$&$0$ & $0$\\
$b_{2j}$ & $BE/2v^2_\sigma$ & $ (CE+BF)\alpha^2/2v^2_\sigma$ & $ CF \alpha^4/4v^2_\sigma$& $0$\\
$a_{3j}$ &$ - (CE+BF)/2v_\sigma$ & $- CF \alpha^2/2v_\sigma $&$0$ &$0$ \\
$b_{3j}$ & $- (CE+BF)\alpha^2/4v^3_\sigma$& $- BE \alpha^2/2v^3_\sigma- CF\alpha^4/8v^3_\sigma$ &$ (CE+BF)\alpha^4/4v^3_\sigma
$ & $ CF \alpha^6/12v^3_\sigma$ \\
$a_{4j}$ &$ -CD-AF$& $0$& $0$&$0$ \\
$b_{4j}$ &$0$ & $ (BD+AE)\alpha^2/2v^2_\sigma$ & $ (CD+AF)\alpha^4/8v^2_\sigma$& $0$\\
$a_{5j}$ &$- (BD+AE)/v_\sigma$ &$ - (CD+AF) \alpha^2/2v_\sigma$ &$0$ &$0$ \\
$b_{5j}$ & $- (BD+AE)\alpha^2/2v^3_\sigma$& $-(CD+AF) \alpha^4/8 v^3_\sigma$ & $ (BD+AE)\alpha^4/8v^3_\sigma$ & $(CD+AF)\alpha^6/48v^3_\sigma$ \\
\hline \hline
\end{tabular}
\caption{\label{tab:Gsn}  Coefficients of the singular expansion (\ref{eq:singexp1}) for the case $XY=sn$.}
\end{center}
\end{table}
\begin{table}[h]
\begin{center}
\begin{tabular}{||c|c|c|c|c||}
\hline \hline
 & $j=0$ & $j=1$ & $j=2$ & $j=3$ \\
\hline 
$a_{1j}$ & $G^2/2v^2_\sigma$ & $0$& $0$& $0$\\
$b_{1j}$ & $G^2 \alpha^2/2v^4_\sigma$& $0$&$0$ &$0$ \\
$a_{2j}$ & $-G^2/2v^2_\sigma$& $0$&$0$ &$0$\\
$b_{2j}$ & $-G^2\alpha^2/2v^4_\sigma$& $0$& $G^2\alpha^4/4v^4_\sigma$& $0$\\
$a_{3j}$ & $0$& $-G^2\alpha^2/2v^3_\sigma$& $0$&$0$ \\
$b_{3j}$ & $0$& $-5G^2\alpha^4/8v^5_\sigma $ & $0$&$G^2\alpha^6/12v^5_\sigma$ \\
\hline \hline
\end{tabular}
\caption{\label{tab:Goo} Same as Table~\ref{tab:Gsn} for the case $XY=\omega \omega$.}
\end{center}
\end{table}
The cases $XY=s \omega, n \omega$ follow a different singular expansion. They generically read:
\begin{eqnarray}
  \nonumber \tilde{G}^{sing}_{XY} (k;\tau,\tau') & = & c_1 k + c_2 k \cos (2 v_{\sigma} L k) + (c_3 k^2 +d_3) \sin (2 v_{\sigma} L k) \\
\label{eq:singexp2} & + & c_4 k \cos ( v_{\sigma} L k) + (c_5 k^2 +d_5) \sin ( v_{\sigma} L k) \ .
\end{eqnarray}
The functions $c_i,d_i$ admit an expansion in powers of $L$:
\be c_i=\sum_j c_{ij} L^j \ ,  \quad d_i= \sum_j d_{ij} L^j \ . \ee
In Table~\ref{tab:Gso} we show the coefficients $c_{ij},d_{ij}$ for the case $XY=s \omega$. Finally, the case $XY=n \omega$ uses the same expansion but
with the change $B \rightarrow E, C \rightarrow F, A \rightarrow D$.
\begin{table}[h]
\begin{center}
\begin{tabular}{||c|c|c|c|c||}
\hline \hline
 & $j=0$ & $j=1$ & $j=2$ & $j=3$ \\
\hline
$c_{1j}$ & $i BG/2v^2_\sigma $& $0$ & $0$ & $0$ \\
$c_{2j}$ & $-iBG/2v^2_\sigma $ & $-iCG \alpha^2/2v^2_\sigma$ & $ 0$& $0$\\
$c_{3j}$ & $iCG/2v_\sigma$ & $0$ & $0 $ & $0$ \\
$d_{3j}$ & $iCG\alpha^2/4v^3_\sigma$ & $-iBG\alpha^2/2v^3_\sigma $ & $-iCG\alpha^4/4v^3_\sigma$ & $0$ \\
$c_{4j}$ & $0$& $-iAG\alpha^2/2v^2_\sigma $ & $0$ & $0$ \\
$c_{5j}$ & $iAG/v_\sigma $ &$0$ & $0$ & $0$ \\
$d_{5j}$ & $iAG\alpha^2/2v^3_\sigma $ & $0$ & $-iAG\alpha^4/8v^3_\sigma $ & $0$ \\
\hline \hline
\end{tabular}
\caption{\label{tab:Gso} Coefficients of the singular expansion (\ref{eq:singexp2}) for the case $XY=s\omega$.}
\end{center}
\end{table}

For the ease of simplicity, in this Appendix we have defined the following quantities:
\begin{eqnarray}
\nonumber A & = & - \frac{\mu}{T} \frac{v_s^2}{v_{\sigma}^2} \ , \quad  B = \alpha \frac{s}{n}  \frac{v^2_\sigma-v^2_n}{v^2_\sigma} \left(
 \frac{\tau'}{\tau} \right)^{\alpha} \ , \quad C  =   \frac{s}{n}  \frac{v^2_\sigma-v^2_n}{v^2_\sigma} \left(
 \frac{\tau'}{\tau} \right)^{\alpha} \ , \quad D =  \frac{v_n^2}{v_{\sigma}^2} \ , \\
\nonumber E & = & \alpha  \frac{v^2_\sigma-v_n^2}{v^2_\sigma} \left(
 \frac{\tau'}{\tau} \right)^{\alpha}  \ , \quad F =  \frac{v^2_\sigma - v^2_n}{v^2_\sigma} \left(
 \frac{\tau'}{\tau} \right)^{\alpha} \ , \quad G =  \frac{s}{n} \left(v^2_\sigma - v^2_n \right) \left( \frac{\tau}{\tau'}\right)^\alpha \ .
\end{eqnarray}


\begin{thebibliography}{99}

\bibitem{Aoki2006a}
Y. Aoki, G. Endrodi, Z. Fodor, S. D. Katz, and K. K. Szabo, Nature {\bf 443}, 675 (2006).

\bibitem{Aoki2006b}
Y. Aoki, Z. Fodor, S. D. Katz, and K. K. Szabo, Phys. Lett. {\bf B643}, 46 (2006). 

\bibitem{Aoki2009}
Y. Aoki, Sz. Borsanyi, S. Durr, Z. Fodor, S. D. Katz, S. Krieg, and K. K. Szabo, JHEP {\bf 0906}, 088 (2009).

\bibitem{Borsanyi2010}
S. Borsanyi, Z. Fodor, C. Hoelbling, S. D. Katz, S. Krieg, C. Ratti, and K. K. Szabo,  JHEP {\bf 1009}, 073 (2010).

\bibitem{Bazavov2012}
A. Bazavov, {\it et al.}, Phys. Rev. D {\bf 85}, 054503 (2012).

\bibitem{asakawa89} M. Asakawa and K. Yazaki, Nucl. Phys. {\bf A504}, 668 (1989).

\bibitem{berges98} J. Berges and K. Rajagopal, Nucl. Phys. {\bf B538}, 215 (1999).

\bibitem{scavenius01}
O. Scavenius, A. M\`ocsy, I. N. Mishustin, and D. H. Rischke, Phys. Rev. C 
{\bf 64}, 045202 (2001).

\bibitem{barducci} A. Barducci, R. Casalbuoni, S. De Curtis, R. Gatto, and G. Pettini, Phys. Lett. {\bf B231}, 463 (1989); Phys. Rev. D {\bf 41}, 1610 (1990); A. Barducci, R. Casalbuoni, G. Pettini, and R. Gatto, {\it ibid.} 
{\bf 49}, 426 (1994).

\bibitem{halasz98} M. A. Halasz, A. D. Jackson, R. E. Shrock, M. A. Stephanov, and J. M. Verbaarschot, Phys. Rev. D {\bf 58}, 096007 (1998).

\bibitem{hatta02} Y. Hatta and T. Ikeda, Phys. Rev. D {\bf 67}, 014028 (2003).

\bibitem{antoniou02} N. G. Antoniou and A. S. Kapoyannis, Phys. Lett. 
{\bf B563}, 165 (2003).

\bibitem{stephanov}
M. Stephanov, Prog. Theor. Phys. Suppl. {\bf 153}, 139 (2004);
Int. J. Mod. Phys. A {\bf 20}, 4387 (2005); PoS(LAT2006)024.

\bibitem{MohantyQM}
B. Mohanty, Nucl. Phys. {\bf A 830}, 899c (2009) . 

\bibitem{Shuryak1998}
M. A. Stephanov, K. Rajagopal, and E. Shuryak, Phys. Rev. Lett. {\bf 81}, 4816 (1998); Phys. Rev. D {\bf 60}, 114028 (1999).

\bibitem{Hatta2003}
Y. Hatta and M. A. Stephanov, Phys. Rev. Lett. {\bf 91}, 102003 (2003); {\bf 91}, 129901(E) (2003).

\bibitem{small}
B. Berdnikov and K. Rajagopal, Phys. Rev. D {\bf 61}, 105017 (2000). 
 
\bibitem{nongaussian}
M. A. Stephanov, Phys. Rev. Lett. {\bf 102}, 032301 (2009).

\bibitem{STAR2010}
M. M. Aggarwal {\it et al.} (STAR Collaboration) Phys. Rev. Lett. {\bf 105}, 022302 (2010).

\bibitem{Kapusta2012}
J. I. Kapusta, B. M\"uller, and M. Stephanov, Phys. Rev. C {\bf 85}, 054906 (2012).

\bibitem{Gavin-Aziz}
  S. Gavin and M. Abdel-Aziz, Phys. Rev. Lett. {\bf 97}, 162302 (2006).

\bibitem{Dobado}
 A. Dobado, F. J. Llanes-Estrada and J. M. Torres-Rincon, Eur. Phys. J. C {\bf 72}, 1873 (2012).

\bibitem{KapustaGale}
J. I. Kapusta and C. Gale,
\emph{Finite Temperature Field Theory},	Cambridge University Press, Cambridge, 2nd edition, 2006.

\bibitem{Kapusta2010}
J. I. Kapusta, Phys. Rev. C {\bf 81}, 055201 (2010).

\bibitem{Bower:2001fq} 
D. Bower and S. Gavin, Phys. Rev. C {\bf 64}, 051902 (2001).

\bibitem{Randrup2009}
J. Randrup, Phys. Rev. C {\bf 79}, 054911 (2009).

\bibitem{Eggers}
J. Eggers, Phys. Rev. Lett. {\bf 89}, 084502 (2002).

\bibitem{Kang}
W. Kang and U. Landman, Phys. Rev. Lett. {\bf 98}, 064504 (2007).

\bibitem{Fixman1962}
M. Fixman, J. Chem. Phys. {\bf 36}, 310 (1962).

\bibitem{Kawaski1970}
K. Kawasaki, Ann. Phys. {\bf 61}, 1 (1970).

\bibitem{Kawaski1976}
K. Kawasaki, in {\it Phase Transitions and Critical Phenomena}, edited by C. Domb and M. S. Green (Academic, New York, 1976), Vol. 5a, p. 165.

\bibitem{Kadanov1968}
L. P. Kadanoff and J. Swift, Phys. Rev. {\bf 165}, 310 (1968).

\bibitem{Zwanzig1972}
R. Zwanzig, in {\it Statistical Mechanics: New Concepts, New Problems, New Applications}, edited by S. A. Rice, K. F. Freed, and J. C. Light (University of Chicago, Chicago, 1972), p. 241.

\bibitem{Sengers1995}
J. Luettmer-Strathmann, J. V. Sengers, and G. A. Olchowy, J. Chem. Phys. {\bf 103}, 7482 (1995).

\bibitem{Hao1991}
Hong Hao, Ph.D. thesis, University of Maryland, College Park, Maryland, 1991.

\bibitem{Berg1988}
R. F. Berg and M. R. Moldover, J. Chem. Phys. {\bf 89}, 3694 (1988).

\bibitem{Nieuwoudt1989}
J. C. Nieuwoudt and J. V. Sengers, J. Chem. Phys. {\bf 90}, 457 (1989).

\bibitem{Rajagopal2000}
K. Rajagopal and F. Wilczek, in {\it At the Frontier of Particle Physics / Handbook of QCD}, edited by M. Shifman (World Scientific, 2001), Vol. 3, p. 2061.

\bibitem{Son2004}
D. T. Son and M. A. Stephanov, Phys. Rev. D {\bf 70}, 056001 (2004).

\bibitem{Hohenberg1977}
P. C. Hohenberg and B. I. Halperin, Rev. Mod. Phys. {\bf 49}, 435 (1977).

\bibitem{AMY2003}
P. Arnold, G. D. Moore, and L. G. Yaffe, JHEP {\bf 0305}, 051 (2003).

\bibitem{Hosoya1985}
A. Hosoya and K. Kajantie, Nucl. Phys. {\bf B250}, 666 (1985).

\bibitem{Gavin1985}
S. Gavin, Nucl. Phys. {\bf A435}, 826 (1985).

\bibitem{Prakash1993}
M. Prakash, M. Prakash, R. Venugopalan, and G. Welke, Phys. Rep. {\bf 227}, 321 (1993).

\bibitem{torresphd} 
  J. M. Torres-Rincon, Ph.D. thesis, Universidad Complutense de Madrid, Spain, 2012
  arXiv:1205.0782 [hep-ph].

\bibitem{FernandezFraile:2005ka} 
  D. Fernandez-Fraile and A. Gomez Nicola,
  Phys. Rev. D {\bf 73}, 045025 (2006).


\bibitem{Skokov2010} 
  V. V. Skokov and D. N. Voskresensky, Nucl. Phys. A {\bf 847}, 253 (2010)
 

\bibitem{Shuryak:2009cy}
  E. Shuryak,
  Phys. Rev. C {\bf 80}, 054908 (2009).
  
\bibitem{Staig:2010pn}
  P. Staig and E. Shuryak,
  Phys. Rev. C {\bf 84}, 034908 (2011).
  
\bibitem{Staig:2011wj}
  P. Staig, E. Shuryak,
  Phys. Rev. C {\bf 84}, 044912 (2011).


\end{thebibliography}
\end{document}